\def\etal{{\it et al. }}
\def\eh{E_H}
\def\D{\Delta_0}
\def\Dk{\Delta_{\bf k}}
\def\wi{\omega_1}
\def\wii{\omega_2}
\def\zk{\zeta_{\bf k}}
\def\kn{{\bf k}_n}
\def\wn{\omega_n}

\def\dop{${\bf v}_s\cdot {\bf k}$}
\def\beq{\begin{equation}}
\def\seq{\end{equation}}
\def\beqs{\begin{eqnarray}}
\def\seqs{\end{eqnarray}}
\def\ei{\epsilon_1}
\def\eii{\epsilon_2}
\def\ep{\epsilon}
\def\eab{E_{ab}}

\def\lb{\label}

\def\w{\omega}

\def\vs{{\bf v}_s}
\def\wt{\widetilde\omega}
\def\vsr{{\bf v}_s ({\bf r})}
\documentstyle[aps,prb,multicol,epsf,rotate]{revtex}
\begin{document}
\draft 
\twocolumn[\hsize\textwidth\columnwidth\hsize\csname@twocolumnfalse\endcsname
\title{Thermodynamics of  $d$-wave superconductors in
a magnetic field}
\author{I. Vekhter$^1$\cite{IV},  P. J. Hirschfeld$^2$, E. J. Nicol$^3$}
\address{${}^1$Theoretical Division, Los Alamos National Laboratory,
	Los Alamos, NM 87545\\
	${}^2$Department of Physics, University of Florida,
	Gainesville, FL 32611\\
	${}^3$Department of Physics, University of Guelph, Guelph, 
	Ontario N1G 2W1, Canada\\
	}
\date{\today}
\maketitle
\begin{abstract}
We investigate the thermodynamic properties of a two-dimensional $d$-wave
superconductor in the vortex state using a semiclassical approach, and
argue 
that such an approach is valid for the analysis of the experimental
data on high-temperature superconductors.
We develop a formalism where the spatial average of a physical quantity
is written as an integral over the probability density of the Doppler
shift, and evaluate this probability density for several model cases.
The approach is then used to analyse the behavior of the specific
heat and the NMR spin-lattice relaxation rate in a magnetic field.
We compare our results with the experimental measurements, and explain the
origin of the discrepancy between the results from different groups.
We also address the observability of the recently predicted fourfold
oscillations of the specific heat for the magnetic field parallel to the
copper oxide planes. We consider both the orbital and the Zeeman effects, and
conclude that at experimentally relevant temperatures Zeeman
splitting does not appreciably reduce the anisotropy, although it does
change the field dependence of the anisotropic specific heat.
We predict a scaling law for the non-exponentially decaying NMR
magnetization, and discuss different approaches to the effective relaxation
rate.
\end{abstract}

%\pacs{74.72.-h, 74.25.Fy, 74.25.Ld}
\twocolumn
\vskip.2pc]

\section{Introduction}

Despite significant recent advances we still lack a complete understanding 
 of the physics of low energy excitations in the vortex state of 
unconventional 
superconductors.
High-temperature superconductors (HTSCs)
are an example of a system where 
theoretical
predictions can be checked against a large bodof experimental evidence.
In zero field these materials have a $d$-wave 
superconducting energy gap, with nodes along the diagonals of the Brillouin 
zone, and consequently a finite density
of  low energy excitations. \cite{vanHarlingen}
Moreover, it is believed that at temperatures low compared to the transition 
temperature, $T\ll T_c$, these excitations 
are reasonably well described
by the Landau quasiparticles, even though such an approach
fails in these materials at higher energies. 
A variety of experimentally measured quantities
such as the electronic specific heat \cite{Moler,Phillips,Revaz,Junod},
effective penetration depth from $\mu$SR \cite{Sonier},
spin-lattice relaxation rate \cite{Slichter,Guo},
and thermal conductivity \cite{Chiao,Aubin,Ong}
are available to test the predictions of theories.

In this work we discuss the influence of the magnetic field on
the thermodynamic quantities in the vortex state of the unconventional
superconductor,  and, in particular, address the question of how these
properties depend on the structure of the vortex state.
 We concentrate on the behavior of the 
density of states and the electronic specific heat and the
spin-lattice relaxation of the NMR magnetization. 
There exist several theoretical approaches to the analysis
of the thermodynamic quantities in the vortex state of 
unconventional superconductors. We employ here a semiclassical 
approach \cite{Volovik,Kuebert1}, which has been successful
in describing the field dependence of a variety of
the physical quantities \cite{Volovik,Kuebert1,Kuebert2}. It is an
approximate description, and in the next section we discuss 
the region of its validity and the grounds for our belief 
that it is applicable
to the present problem. Section \ref{sec:semiclassics} introduces
the basic model of the nodal quasiparticles and the idea that 
the physical quantities in the vortex state can be obtained 
by calculating the spatial average of their local values, computed
with the help of
the semiclassical approach.

Until now such spatial averages have only been done analytically in 
an oversimplified model of a single vortex.
\cite{Volovik,Kuebert1,Kuebert2,Vekhter1,Vekhter2,Barash} 
In this paper we introduce a generalization of
these approaches by rewriting the spatial
average as an average over the probability density of
the Doppler shift
of the quasiparticle energy in the presence of the superflow.
 Restating the problem in this language 
enables us to introduce several model
distributions of the probability density, discussed in
Section \ref{sec:distributions},
 and investigate how 
the physical quantities obtained within the semiclassical framework
depend on these distributions and on the structure of the vortex state.
We obtain the
energy and field dependence of the density of states for the geometries with
the magnetic field applied both normal to the superconducting planes
and in the plane in Section \ref{sec:dos}. This density of states
is used to analyze the behavior of the electronic specific heat in HTSCs.
We obtain the energy scales relevant to the high-temperature
superconductors in the vortex state, and suggest
a resolution to the origin of the disagreement between
different experimental groups regarding the magnitude of the
field-dependent term in the specific heat and the form of the scaling
function; this is the content of Section \ref{sec:spheat}. In the same
Section we address the question of the observability
of the oscillations in the specific heat for the magnetic field applied in the
superconducting plane as a function of the angle between the field
and the nodal directions. These oscillations have been recently predicted
\cite{Vekhter2,volovik2}, but so far have not been observed.
Part of the difficulty may stem from the smallness of the
in-plane Doppler energy scale, 
as inferred from the experimental measurements; it has recently been
argued that the Zeeman splitting reduces the observed oscillations
significantly,\cite{Whelan} 
and we investigate its effect in detail in this work.

Section \ref{sec:nmr} is devoted to the effect of the nonuniform
density of states on the spin-lattice relaxation time. This 
non-uniformity leads to a non-exponential decay of the magnetization
and to a field dependence of the effective relaxation rate\cite{Vekhter1};
here we show that the effective relaxation rate depends on the structure of
the vortex state, and obtain an approximate form for it. We also
predict a scaling law for the magnetization decay which
can be checked directly. The effect of impurities on
the density of states is briefly addressed in Section \ref{sec:impurity}.
We expect these effects to be very important for the discussion of
the transport properties which we defer to a later publication.
Finally we summarize our findings and discuss some open questions.

\section{Semiclassical approximation}
\label{sec:validity}

HTSCs are extreme type II superconductors
 (the ratio of the London penetration depth $\lambda_L$ to the 
coherence length $\xi_0$ is large, $\lambda_L/\xi_0\sim 100$), 
and are in the mixed state over the 
range of applied fields, $H$,
 from a few hundred Gauss to well in excess of
50 Tesla in YBa$_2$Cu$_3$O$_{7-\delta}$ (YBCO)
and Bi$_2$Sr$_2$CaCu$_2$O$_{8+\delta}$ (Bi-2212) near optimal doping.
In the mixed states the magnetic field penetrates the bulk of the
superconductor in the form of vortices, which consist of the 
cores, where the superconducting 
order parameter is suppressed, and circulating supercurrents around them.
The vortex core size is of the order
of the coherence length, $\xi_0\sim 15$\AA \cite{Fischer,Davis}, 
while the average intervortex 
distance can be estimated
by imposing the requirement of one flux 
quantum $\Phi_0=hc/2e$ per vortex, or
$d/2=R=\sqrt{\Phi_0/\pi B}$, where $B$ is the internal field.
At typical experimentally accessible fields (1-20 T) 
$\lambda_L\gg d\gg \xi_0$, 
the magnetization due to the
vortex lattice is small, and the internal field can be 
replaced by the applied field, $H$, so that
$d\sqrt{H}\sim 500$\AA$\cdot$T$^{1/2}$.
The actual distance differs
from the average value $d$ by a numerical factor of the order of unity, 
which depends
 on the structure of the vortex state; the vortices in HTSCs may
         form a regular lattice (as they do in YBCO and in
Bi-2212 at low fields \cite{Fischer-YBCO,SANS}) or be moderately
disordered
(as in Bi-2212 at higher fields \cite{Davis}). 

In the experimentally relevant fields, $H_{c1}\ll H\ll H_{c2}$, where 
$H_{c1}$ ($H_{c2}$) is the lower (upper) critical field, there may exist two
types of low energy excitations. First, as in conventional, $s$-wave
materials with an isotropic gap, 
there may be a branch of 
low-energy fermionic excitations bound to
the vortex cores.\cite{deGennes} Experimental evidence 
suggests, however, that there
is at most one such  state in the vortex cores of YBCO and Bi-2212
\cite{Fischer-YBCO,Davis}, and therefore the properties of
the mixed state of $d$-wave superconductors are dominated by the ``extended''
quasiparticle states in the bulk.
These states are formed when 
quasiparticles with momenta close to the position of the 
nodes of the gap, $\Dk$, in the momentum space (and therefore 
with a small gap) 
interact with the supercurrents in the vortex state. 
Most of the theoretical work has therefore  
explored the properties of these states.

Very significant progress has been made by utilizing the semiclassical 
approach, which treats the momentum and the position 
of the quasiparticle as commuting variables. It is 
valid when the wave function of a quasiparticle can be replaced
by its envelope on the length scales exceeding the coherence length,
i.e. when $k_f\xi_0\gg 1$, where
$k_f$ is the inverse Fermi wavelength.
In that method
the effect of the supercurrents is accounted for by 
introducing a Doppler shift into
the quasiparticle energy spectrum, \cite{Tsuneto,Yip,Volovik,Kuebert1}
$E^{\prime} ({\bf k}, {\bf r})=E({\bf k})+ \epsilon ({\bf k}, {\bf r})$. 
Here $E({\bf k})$ 
is the energy of a quasiparticle
with  momentum {\bf k} in the absence of the  field measured with respect to
the chemical potential.
In a two-dimensional $d$-wave superconductor this spectrum is
conical
(massless anisotropic Dirac spectrum): 
 $E({\bf k})\approx\pm\sqrt{v_f^2 k_\perp^2 +v_\Delta^2 k_{\|}^2}$, 
where the Fermi velocity $v_f$ is 
associated with the dispersion of the quasiparticles in the direction normal
to the Fermi surface (component $k_\perp$ of the momentum), 
while $v_\Delta\sim\Delta_0/k_f$ is the slope of
the gap at the node associated with the dispersion of the quasiparticles along
the Fermi surface ($k_{\|}$).
The Doppler shift, 
$\epsilon ({\bf k}, {\bf r})=$\dop { }depends on the quasiparticle momentum 
and the local value of the supervelocity, $\vs ({\bf r})$. 
This shift in the energy is an exact result for a uniform supercurrent,
\cite{Tinkham} where it reflects the pairing of the electrons
with a finite center of mass momentum. 
In the simplest picture such an approach remains 
valid
for a non-uniform current 
for as long as  the spatial
variations of $\vs$ are slow on the scale of the spatial extent
of the Cooper pair, $\xi_0$.
In superconductors with nodes in the energy gap, the Doppler shift
may exceed the local (in the momentum space) gap, and leads to
an increase in the density of the unpaired quasiparticles: 
even at $T=0$
for some positive energies $E$ the shifted energy $E^\prime$ is
negative so that the corresponding states become occupied.
In the context of $d$-wave superconductors this was emphasized
by Yip and Sauls\cite{Yip}, who investigated the effect of
the screening currents in the Meissner state 
on the superfluid density.\cite{Yip} These currents vary
on the scale of the penetration depth $\lambda_L\gg\xi_0$, 
so that the Doppler shift description is appropriate, and 
result in  a linear dependence of
the effective
penetration depth on the applied field for certain experimental 
geometries \cite{Yip}. However so far
the predicted dependence  has not been confirmed
experimentally.\cite{Bidinosti}

Similar physics is at play in the dilute vortex limit. 
At distances small compared to the penetration depth,
supercurrents around  an isolated vortex 
are inversely proportional to the 
distance from the center of the vortex, $r$; for $\xi_0\ll r\ll \lambda_L$
the supervelocity field is
 $|\vs|=\hbar/2mr$, where $m$ is the quasiparticle mass. 
Consequently, the requirement of the slowness of the variation of $\vs$,
which can be written as $|\nabla \vs|\xi_0<< \vs$ (two particles comprising
the Cooper pair ``see'' the same velocity), is satisfied at $r\gg\xi_0$, 
justifying the use of the semiclassical approach outside of the core and 
therefore for the analysis of the extended quasiparticle excitations
at energies low compared to the gap maximum.
Such an analysis for a single vortex in
$d$-wave superconductors  was first carried out 
in a landmark paper by Volovik\cite{Volovik},
who showed that the density of the extended quasiparticles 
at low temperature, $T$, varies
as $\sqrt H$. 
This result was confirmed first by Moler {\it et al.} 
\cite{Moler} and 
subsequently by other groups \cite{Phillips,Revaz,Junod} from the
measurements of the electronic specific heat in an applied magnetic field.
Numerical studies of the tight-binding model are also in qualitative agreement
with this result.\cite{Wang}
Moreover Volovik has shown that the density of the extended quasiparticles
dominates that of the states bound to the vortex core even if the latter
set is treated as a quasi-continuum \cite{Volovik} (as it would be in a 
superconductor with a long coherence length), providing 
further theoretical support for neglecting the
core states in the analysis of the properties of the vortex state in 
unconventional superconductors. 

The semiclassical approach was
incorporated into the Green's function formalism by K\"ubert and Hirschfeld
\cite{Kuebert1}, and used in that form to analyze 
thermodynamic and transport properties of the high-T$_c$ cuprates in the
vortex state.\cite{Kuebert2,Vekhter1,Vekhter2} In particular, 
accounting for the impurity scattering in this framework has
significantly improved  the agreement between the theory and
the measurements of the electronic specific heat \cite{Kuebert1}, and
the  field dependence of the low-temperature thermal conductivity
\cite{Chiao} is in qualitative agreement with the results of a semiclassical 
calculation \cite{Kuebert2}. In the semiclassical
 approach the 
effect of the magnetic field is contained in 
the new energy scale associated with
the Doppler shift, $E_H=v_f/d$, and the behavior of the physical properties is
determined by the competition between this energy, the temperature, and the
impurity scattering rate.
Photoemission measurements on high-T$_c$ compounds 
suggest \cite{ARPES,Mesot} 
that $v_f\simeq 1.5-2.5\times 10^7$ cm/s leading to
 $E_H\sim 30 \sqrt H$ K$\cdot$T$^{-1/2}$.

For a long time, understanding of the low-energy 
excitations in the vortex state beyond this semiclassical picture
has proved elusive. 
The difficulties stem in part from the need to treat on equal footing the
applied magnetic field and the superconducting currents (semiclassical
approach treats the supercurrents classically).
Attempts have been made to take as a starting point
the Landau quantization
of the quasiparticle states, and include the
effects of supercurrents perturbatively,\cite{Gorkov,Anderson}
however, since the supervelocity field is 
long-ranged and singular at the position of each vortex
the Landau levels are strongly mixed, making a detailed
analysis difficult.\cite{Melnikov}
 
The most significant progress has been made in a recent work by
Franz and Tesanovic,\cite{FT} who have introduced a gauge transformation 
which takes into account both the supercurrent distribution and the
magnetic field. In their approach the problem is mapped onto that of
nodal Dirac fermions in an effective zero average magnetic field
interacting with effective scalar and
vector potentials which are periodic in the unit cell of the vortex lattice.
Both Franz and Tesanovic \cite{FT}
and Marinelli {\it et al.}\cite{Marinelli} have studied the band structure
of the nodal quasiparticles for perfectly periodic vortex lattices
for various values of the anisotropy of the Dirac spectrum,
$\alpha_D=v_f/v_\Delta$. 

There are two reasons for expecting modifications to the
semiclassical spectrum. The first is related to the 
singular spatial structure of the supervelocity field. One flux 
quantum associated
with a single vortex means that the superconducting order parameter, or,
equivalently, the wave function of a Cooper pair (charge $2e$)
is single valued and  
has a phase winding
of $2\pi$ around each vortex. As emphasized in Ref.\onlinecite{FT},
the semiclassical approach transfers this phase winding equally
to each of the 
quasiparticles forming the Cooper pair (each having charge $e$).
Consequently, their
wave functions change phase by $\pi$ around a vortex line 
(Aharonov-Bohm phase), 
leading
to the necessity of introducing branch cuts and the problem of
multi-valued wave functions in the full quantum mechanical treatment.
However, the semiclassical approximation is only valid for
large quantum numbers, that is for the quasiparticles for which
the total phase of the wave function, accumulated
as the electron moves around the vortex, is large.
The wave function of an electron
circling a vortex  at a distance $r$ from the vortex center acquires a 
phase $2\pi k_f r$, compared to an extra Aharonov-Bohm
phase $\pi$ from the supervelocity
field. 
For the analysis of the extended states ($r\gg\xi_0$) in
the semiclassical approach (valid at $k_f\xi_0\gg 1$),
$ k_f r\gg 1$, so that if the phase of the wave function
is changed by $\pi$, it still 
corresponds to the quasiclassical state with essentially the same
energy and momentum.
Since
$\xi_0\sim v_f/\Delta_0$, we can rewrite
the condition for the applicability
of the semiclassical method $k_f\xi_0\gg 1$
as $\alpha_D=v_f/v_\Delta\gg 1$. 
Indeed, the work of Refs.\onlinecite{FT,Marinelli} has shown that
for large anisotropy of the Dirac cone the semiclassical approach remains
valid down to the lowest energies. 
Since $\alpha\simeq 14$ for YBCO \cite{Chiao},
and $\alpha\simeq 20$ for Bi-2212 \cite{Chiao2,Mesot}
this is the parameter range relevant for the study of HTSCs.
In a very recent preprint Mel'nikov has shown that the 
Aharonov-Bohm phase leads to a different result for the quasiparticle density
at distances $r \gg \lambda_L$, while in the range 
$\xi_0 \ll r\ll\lambda_L$
the semiclassical results hold.\cite{Melnikov2} 
Once again, since in the field range where most experimental measurements
are done the intervortex distance $d\ll\lambda_L$ this result suggests that
the semiclassical approach is adequate for the analysis of these
experiments.

Quantum mechanical treatment is nevertheless needed for 
accurate description of the states at very low energies. 
Kopnin and Volovik \cite{Volovik,Kopnin} have considered the effect of
the magnetic field on the nodal quasiparticles perturbatively, and found that
the spacing between quantum mechanical levels of the near nodal quasiparticles
for which the spatial extent of the wave function is comparable to the
intervortex distance is $E_{KV}=v_\Delta/d=E_H/\alpha_D$. 
Therefore they have argued that
below this energy scale the semiclassical approach becomes invalid.
For $\alpha \sim 15$
this energy scale is of the order of a few Kelvin per square root of Tesla.
However the specific heat measurements show no 
crossover to a novel behavior at that scale\cite{Junod}, and the 
measurements of the thermal conductivity below 0.5 K in fields of up
to 8 Tesla are in agreement with the semiclassical calculations.\cite{Chiao}

Marinelli {\it et al.} \cite{Marinelli}
have suggested on the basis of their numerical work
that the actual crossover energy decreases much more rapidly with
the increasing anisotropy $\alpha$ than the result of Ref.\onlinecite{Kopnin}
implies.
They have noted that the dispersion of the quasiparticles along the 
Fermi surface is strongly softened  in the vortex state, so that
the effective anisotropy \cite{Marinelli}
$\alpha_R$ increases much more rapidly than
$\alpha_D$, leading to a much smaller crossover scale, perhaps exponentially
small in $\alpha_D$.\cite{Simon1} 
Since in real samples
the presence of impurity scattering and the disorder in the vortex lattice
always smear out the energy structure on small scales, we therefore expect 
that for the purposes of comparison with the measurements of
the thermodynamic quantities, the semiclassical 
description is adequate.

Therefore for the parameter range relevant to the study of most
real unconventional superconductors,
the semiclassical approach reproduces the
energy spectrum of the near-nodal quasiparticles in the
vortex state to a high degree of accuracy.
 Moreover, presently it remains the only
approach which is capable of including the effect of impurity scattering 
into the analysis, and we use it hereafter.

\section{Semiclassical approach to the vortex state}
\label{sec:semiclassics}

\subsection{Nodal approximation}

The semiclassical approximation takes as its starting point a Fermi-liquid
description of the nodal quasiparticles, so that in 
the absence of a magnetic field
the Green's function
in the particle-hole (Nambu) space is given by 
the Bardeen-Cooper-Schrieffer form with an 
anisotropic gap,
\beq
\lb{BCSGreen}
G({\bf k},\omega_n)=-
\frac
{i\omega_n\widehat\tau_0+\Delta_{\bf k}\widehat\tau_1+
\zeta_{\bf k}\widehat\tau_3}
{\omega_n^2+\zeta_{\bf k}^2+\Delta_{\bf k}^2}.
\end{equation}
Here $\widehat\tau_i$ for $i=0\ldots 3$ are the Pauli matrices 
($\widehat\tau_0$ is the unit matrix), 
$\omega_n=\pi T (2n+1)$ is the Matsubara frequency, and 
$\zeta_{\bf k}$ is the energy of a quasiparticle with momentum {\bf k}
measured relative to the chemical potential.
We consider a two-dimensional Fermi surface with 
an energy gap of  $d_{x^2-y^2}$ symmetry given 
by $\Dk=\D (k_x^2-k_y^2)/k^2$. 
Low energy properties depend only on the nodal quasiparticles, and
are only functions of the parameters entering
the linearized dispersion near nodes at position $\kn$.
As $\zk\approx {\bf v}_f\cdot
({\bf k}- \kn)$, and
$\Dk\approx {\bf v}_\Delta \cdot
({\bf k}- \kn)$ near a node, the poles of the Green's function 
after analytic continuation to the real axis, 
$i\wn \rightarrow \omega+i\delta$, are located at
energies 
\beq
E({\bf k})=\pm\sqrt{\zk^2+\Dk^2}
\approx\pm\sqrt{v_f^2 k_\perp^2 +v_\Delta^2 k_{\|}^2},
\seq
where $k_\perp$ and $k_{\|}$ are the components of ${\bf k}- \kn$
normal to and along the Fermi surface respectively.
We parameterize the Fermi surface near 
each of the four nodes not by the momenta $k_\perp$ and $k_{\|}$, but by the
quasiparticle energy $E$ and the angle $\Theta$ defined as
\beqs
v_f k_\perp&=&E\sin\Theta,
\\
v_\Delta k_{\|}&=&E\cos\Theta. 
\seqs
The energy cutoff is chosen to preserve the volume of the Brillouin 
Zone, so that for a square lattice with the periodicity $a$, 
it is set at 
$E_0=\sqrt{\pi v_f v_\Delta}/a$.\cite{Durst} 
By making this choice we extend the conical dispersion law 
beyond the maximal gap value, $\Delta_0$. This leads to logarithmic, 
in $E_0/\Delta_0$, corrections to the quantities which depend on the 
cutoff energy.
Since $v_\Delta\sim \Delta_0/k_f$ and $k_f\sim \pi/a$, we obtain
$E_0\sim\sqrt{E_f\Delta_0}$, where $E_f$ is the Fermi energy. 
In the high-T$_c$ materials $E_f\sim 3-10 \Delta_0$, and, 
consequently, the choice of $E_0$ as the cutoff energy does not affect the
results significantly.
Therefore near each node the Green's function at real frequencies can be
written as
\beq
\lb{Green_nodal}
\widehat G(E,\theta;\omega)=
\frac{\wt\widehat\tau_0+E\cos\Theta\widehat\tau_1+
E\sin\Theta\widehat\tau_3}
{\wt^2-E^2}.
\end{equation}

In writing Eq.(\ref{Green_nodal}) we have replaced 
the bare frequency $\omega$ by the renormalized frequency
$\wt$ to include the effect of impurity scattering. We account for 
isotropic strong (phase shift $\pi/2$)	
impurity scattering in the framework of a self-consistent 
$T$-matrix approximation, and consider a particle-hole symmetric system, so
that the only non-vanishing component of the self energy is
proportional to  $\widehat\tau_0$.\cite{Hirschfeld}
Therefore the effect of impurities is to replace
in the Green's function $\omega$ by 
its renormalized value, $\wt=\omega-\Sigma(\wt)$, with
the self-consistency condition 
\beq
\lb{Sigma}
\Sigma(\wt)=-{n_i}\biggl[\sum_{\bf k}G_{11}(\wt)\biggr]^{-1},
\seq
where $n_i$ is the impurity concentration.
In the nodal approximation the integral over the Brillouin Zone can be
written as a sum over the nodal regions
\beq
\sum_{\bf k}G_{11}=\sum_{nodes}\frac{1}{v_f v_\Delta}
\int_0^{E_0}\frac{EdE}{2\pi}\int_0^{2\pi}\frac{d\Theta}{2\pi} 
\frac{\wt}{\wt^2-E^2}.
\seq
Writing $\wt=\wi+i \wii$, we obtain for $E_0\gg |\wt|$
\beq
\lb{G11}
\sum_{\bf k} G_{11}=-\frac{2}{\pi}\frac{\wi+i\wii}{v_f v_\Delta}
\Biggl[
\ln\frac{E_0}{\sqrt{\wi^2+\wii^2}} +i \arctan\frac{\wi}{\wii}\Biggr].
\seq
The well-known relationships for 
the density of states in the pure limit ($\wii=0,\wi=\omega$), and for 
the residual density of states in the presence of impurities 
($\wi=0,\wii=\gamma$) follow easily (cf. Ref.\onlinecite{Durst}) from
\beq 
N(\w) = -\frac{1}{\pi} \sum_{\bf k} {\rm Im }G_{11} ({\bf k},\omega),
\seq
to give 
\beqs
\lb{dospure}
N(\w)&=&\frac{|\omega|}{\pi v_f v_\Delta}, 
\ \ \ \ \ \ \ \ \ \ \ \ \ \ \ \mbox{pure limit}
\\
N(0)&=&\frac{2}{\pi^2}\frac{\gamma}{v_f v_\Delta}\ln\frac{E_0}{\gamma},
\ \ \ \ \ \ \mbox{unitarity}.
\seqs
The self-consistency condition $\wi+i\wii=\w-\Sigma(\wt)$ for the latter 
case is (cf. Ref. \onlinecite{Hirschfeld2})
\beq
\gamma^2=\frac{\pi}{2} n_i v_f v_\Delta 
\Biggl[\ln\frac{E_0}{\gamma}\Biggr]^{-1}.
\seq

\subsection{Doppler shift}

In the semiclassical approach 
to the  the vortex state the presence of a superflow is accounted for by 
introducing the Doppler shift into  the energy
$\w\rightarrow\w+\epsilon({\bf k}, {\bf r})$ \cite{Volovik,Kuebert1}, where
\beq
\epsilon({\bf k}, {\bf r})=\vsr\cdot {\bf k},
\seq
and $\vsr$ is the supervelocity field at a position
{\bf r} due to all vortices.
It was demonstrated by K\"ubert and Hirschfeld \cite{Kuebert1}
that to very high  accuracy the Doppler shift at the node
${\bf k}_n$ can be used to approximate the Doppler shift for the entire
nodal region. Therefore the Green's function near 
each node can be written as
\beq
\lb{Green_doppler}
\widehat g(E,\Theta;\omega; {\bf r})=
\widehat G(E,\Theta;\omega+\epsilon_n({\bf r})),
\end{equation}
where $\widehat G$ is given by Eq.(\ref{Green_nodal}),
$n$ labels the nodes, and 
$\epsilon_n({\bf r})=\vsr\cdot {\bf k}_n$.
For a $d$-wave superconductor there are two pairs of nodes such that
${\bf k}_1=-{\bf k}_3$ and ${\bf k}_2=-{\bf k}_4$, 
so that the possible values for the Doppler shift are
$\pm\epsilon_1$ and
$\pm\epsilon_2$.

In principle now all the physical quantities can be computed with the help
of this Green's function. 
Several comments have to be made
about the assumptions implicitly present in such calculations. 
First,  
we neglect all inelastic processes.
Second, 
there is no additional quasiparticle damping
due to the presence of the vortices: in the absence of impurities
 the lifetime of a quasiparticle
defined by the pole of the Green's function in Eq. (\ref{Green_doppler})
is infinite. The scattering of the nodal quasiparticles by vortices
depends strongly on the nature of the vortex cores; in the 
high-T$_c$ materials this is an unresolved problem \cite{Lee,Franz_vortex}.
If the point of view is taken that the core is identical to that
of a BCS-like superconductor, the neglect of
vortex scattering is reasonable for a 
 vortex lattice with long range order, and it
remains valid when only short range order exists
provided that the lifetime is restricted by the
impurity scattering rather than vortex disorder.
The role of disorder in the vortex lattice is especially important
for transport properties, where there is a competition
between the increase in the density of quasiparticles
and the change in the transport lifetime in an applied field;
several authors have analyzed
its consequences, \cite{Kuebert2,Vekhter_m2s,Vekhter_qc}
and some questions remain
unresolved \cite{Franz1,Vekhter_qc,Ye}.
STM measurements suggest that the vortex lattice is
ordered in YBCO\cite{Fischer-YBCO}, and that short range order
is present in Bi-2212\cite{Davis},
therefore in that case the assumption is justified.
We note that thermodynamic quantities, such as the density of 
states (DOS), depend on a single energy scale,
$E_H=v_f/d$ even in a disordered vortex state
in absence of strong pinning, 
and this dependence appears to be nearly identical for 
the ordered and disordered vortex lattices\cite{Franz1}.
Therefore we expect that the results obtained within 
the semiclassical approach remain at least
qualitatively valid even for a strongly disordered lattice.

Third, in the analysis
of the impurity scattering 
this approach assumes that the positions of vortices 
and of impurities are uncorrelated.
The self energy given by Eq.(\ref{Sigma}) is obtained after averaging
over 
the positions of impurities, and solving this equation
with the Doppler shift included in the
Green's function implies that  the impurity average and the spatial
average are taken independently.

Finally, in the discussion so far we have neglected the 
Zeeman splitting altogether; this is justified when the 
Doppler energy scale exceeds the Zeeman shift.
In the absence of spin-orbit coupling the Zeeman shift is
$\mu H\approx 0.67H$ K$\cdot$ T$^{-1}$, while
the Doppler shift is $E_H\simeq 30\sqrt H$ K$\cdot$ T$^{-1/2}$;
consequently the two become comparable only 
at $H_{cross}\sim 10^3$T, and hence the Zeeman splitting is irrelevant.
On the other hand, for the field applied in the plane, the 
coefficient in the Doppler shift is much reduced \cite{Vekhter2}, and
the Zeeman splitting
is relevant for some experimental geometries.\cite{Whelan} 
Consequently, we will revisit this question 
in the analysis for this configuration.

If we know how to express a physical quantity $F$ in terms of 
the Green's function
we can now compute its local value $F({\bf r})$
 with the local Green's function given by
Eq.(\ref{Green_doppler}). 
We then approximate the field-dependent measured value $F(H)$ by
the spatial average of $F({\bf r})$ \cite{Kuebert1,Kuebert2} 
\beq
\lb{average_space}
F(H)=\frac{1}{A}\int d^2{\bf r} F(\ei({\bf r}), \eii({\bf r})),
\seq
where the integral is taken over the part
of  a unit cell of the vortex lattice
(with the area $A$) in real space
where
the Doppler shift is much smaller than the
gap maximum. Therefore the integration is to be cut off
at distances of the order of $\xi_0$ from the center of each
vortex. In practice
in many cases the contribution of the core region ($r\le\xi_0$) is
small due to the geometric effect (integrals are weighted with
the surface area  $rdr$) 
and the integral can be extended to the entire unit cell.
We note that the averaging procedure is often non-trivial for 
response functions; for the thermal conductivity $\kappa$, for example, 
$\kappa({\bf r})$ or $1/\kappa({\bf r})$ are averaged depending on
the relative orientation of the magnetic field and the 
heat current.\cite{Kuebert2,Vekhter_m2s} 
The average in Eq.(\ref{average_space}) depends on the
distribution of vortices. In practice, this spatial average 
has been  computed analytically only
for the supervelocity field corresponding to
an isolated flux line, cut off at the average intervortex distance,
\cite{Volovik,Kuebert1,Kuebert2,Vekhter1,Vekhter2}
and numerically for the pancake liquid state.\cite{Vekhter_qc}

The starting point of our approach, which simplifies calculations and 
makes possible a generalization of the semiclassical method to an arbitrary 
configuration of vortices is
to rewrite the average as the integral over the
{\it probability distribution} of the Doppler shift for a 
particular
vortex configuration. 
There are, in general, two types of local quantities,
and therefore of averaging procedures, which are required.
The density of states in the absence of impurity scattering, for example, 
is a direct sum of the contributions from each node,
\beqs
\lb{dos}
&&N(\w, {\bf r})= -\frac{1}{2\pi} {\rm Im}\Biggl\{\sum_{\bf k} {\rm Tr }
\widehat G ({\bf k},\omega)\Biggr\}
\\
\nonumber
&&\approx -\frac{1}{2\pi}{\rm Im}\Biggl\{
\sum_{\alpha=\pm\atop n=1,2}
\int \frac{dE d\Theta}{4\pi^2v_f v_\Delta}
{\rm Tr }
\widehat G(E,\Theta;\omega+\alpha\epsilon_n({\bf r}))\Biggr\},
\seqs
and can consequently be expressed as an integral over the
probability density of the Doppler shift {\it at a single node},
\beq
\lb{dospurefield}
N(\w,H)=\frac{1}{2}\sum_{\alpha=\pm} 
\int_{-\infty}^{+\infty} d\ep N(\w+\alpha\ep){\cal P}(\epsilon),
\seq
where
\beq
\lb{1dist}
{\cal P}(\epsilon)=\frac{1}{A}\int d^2 {\bf r}
\delta\biggl( \epsilon- \vsr\cdot{\bf k}_n\biggr).
\seq 
Such an approach has been recently used to
analyze the behavior of the interlayer conductivity in
the vortex liquid state\cite{Vekhter_qc}, where the function ${\cal P}$
was determined from numerical simulations. A similar method 
(although with an unrealistic distribution, see below) 
has been used
in the analysis of the thermal conductivity.\cite{Yu,Franz1}.

However, in general the function $F$ 
depends on the values for the Doppler shift at two 
inequivalent nodes, $\ei$ and $\eii$,
and the corresponding average can be written as
\beqs
F(H)&=&\int_{-\infty}^{+\infty} d\ei d\eii
F(\ei,\eii) {\cal L}(\epsilon_1, \epsilon_2),
\\
\lb{2dist}
{\cal L}(\epsilon_1, \epsilon_2)&=&
\frac{1}{A}\int d^2 {\bf r}
\delta\biggl( \epsilon_1- \vsr\cdot{\bf k}_1\biggr)
\\ 
\nonumber
&&
\qquad\quad
\times
\delta\biggl(\epsilon_2-\vsr\cdot{\bf k}_2\biggr),
\seqs
where ${\bf k}_1$ and ${\bf k}_2$ label two nearest nodes.
This is the case, for example, for the density of states in
the presence of impurity scattering, since the self energy (implicitly present
in the Green's function in Eq. (\ref{dos})) 
contains the sum over the nodes, see Eq.(\ref{Sigma}), and therefore 
depends on both $\ei$ and $\eii$. In general, the function ${\cal L}$ 
has to be even in both $\ei$ and $\eii$, and symmetric under the interchange
$\ei \leftrightarrow \eii$; in all the cases considered below it
depends on a 
single variable $\ei^2+\eii^2$.

Now all the  relevant
information about the structure of the vortex state 
is contained in the functions 
${\cal L}(\epsilon_1, \epsilon_2)={\cal L}^\prime (\ei^2+\eii^2)$ and
\beq
{\cal P}(\epsilon)= \int d\epsilon_1 {\cal L}(\epsilon, \epsilon_1),
\seq
and therefore to analyze the field dependence of the physical quantities 
we first focus on determining these probability densities.

\section{Probability density for the Doppler shift}
\label{sec:distributions}

The distributions ${\cal P}$ and ${\cal L}$ can be determined numerically
for an arbitrary configuration of vortices. Here we are interested in
making progress analytically, and therefore consider several model
configurations for which the distributions can be found exactly.
Moreover, we propose that the distributions which we consider
give the maximal and the minimal possible weight to the low-energy
Doppler shift, and therefore can be used to obtain the upper and the lower
limits of the experimentally accessible quantities. 

\subsection{Single vortex, ${\bf H}\| \widehat{\bf c}$}

The simplest of these models is that of a velocity field of an 
isolated vortex, cut off at the distance equal to the intervortex distance;
since the experiments are in the dilute vortex limit such an approach
gives an adequate description of the vortex state.
The supervelocity is
$\vsr=\hbar\widehat\theta/2mr$, where $\theta$ is the winding 
angle of the vortex in real space, $m$ is the effective mass, and $r$ is
the distance from the center of the vortex. We now write
the Doppler shift in terms of the energy scale $E_H=v_f/(2R)$, where
$R$ is the radius of the unit cell of the vortex lattice,
taken to be circular,  $R=\sqrt{\Phi_0/\pi H}$, \cite{Kuebert1,Vekhter1}
\beq
\lb{Doppler}
\vsr\cdot {\bf k}_f=\frac{\hbar k_f}{2mr}\sin\theta=\frac{E_H}{\rho}\sin\theta.
\seq
Here we have introduced the normalized length $\rho\equiv{r/ R}$, and have
chosen, without loss of generality, ${\bf k}_n$ along the
direction $\theta=\pi/2$.

The probability distribution at a single node is now easily obtained from
Eq.(\ref{1dist})
\beqs
\lb{Pint}
{\cal P}(\epsilon)&=&{1\over \pi}\int_0^{2\pi} d\theta \int_0^1 \rho d\rho
\delta\biggl( \epsilon- {E_H\over \rho}\sin\theta \biggr)
\\
\nonumber
&=&
{E_H^2\over \pi\epsilon^3} \int_0^{2\pi} d\theta \sin^2\theta
\Theta\biggl({{E_H\over \epsilon}\sin\theta}\biggr)
\Theta\biggl(1-{{E_H\over \epsilon}\sin\theta}\biggr),
\seqs
yielding
\beq
\lb{Psingle}
{\cal P}(\epsilon)=
\cases{ \frac{1}{2}\frac{E_H^2}{\epsilon^3}, & if $\epsilon \ge E_H$;\cr
	\frac{1}{\pi}\frac{E_H^2}{\epsilon^3}
	\biggl[\arcsin\frac{\epsilon}{E_H} - \frac{\epsilon}{E_H}
	\sqrt{1-\frac{\epsilon^2}{E_H^2}} \biggr], & if $\epsilon < E_H$.\cr}
\seq
Here we have taken $\epsilon\ge 0$, the probability density
is even in $\ep$.

It was argued in Ref.\onlinecite{Vekhter_qc} that the function 
${\cal P}(\ep)$ for any vortex configuration
has two important properties. First,
the asymptotic behavior ${\cal P}(\epsilon)=E_H^2/(2\epsilon^3)$ holds
for $\D\gg \ep\gg \eh$. Since the vortices repel each other,
the vortex cores do not overlap. The large Doppler shifts
come from the regions near the cores, where the superfluid velocity is 
high, and consequently are dominated by the single vortex physics.
Second, in the absence of
strong pinning ${\cal P}(\epsilon)$ has a single energy scale
$E_H$ and depends on the Doppler shift only via 
$\ep/E_H$. Since the probability density
is normalized,
\beq
\int_{-\infty}^{+\infty} {\cal P} (\epsilon) d\epsilon =1,
\seq
we can follow Ref.\onlinecite{Vekhter_qc} and define a normalized
dimensionless probability density as
\beq
P(x)=E_H {\cal P}(\epsilon/E_H),
\seq
where $x=\ep/E_H$. 

The two-node probability distribution function is
\beqs
\lb{Ls}
{\cal L}(\epsilon_1, \epsilon_2)=
\frac{1}{\pi}\int_0^{2\pi} d\theta \int_0^1 \rho d\rho
&&\delta\biggl( \epsilon_1- {E_H\over \rho}\sin\theta \biggr)
\\
\nonumber
&&\times
\delta\biggl(\epsilon_2-
\frac{\epsilon_1\sin(\theta+\phi_0)}{\sin\theta}\biggr),
\seqs
where $\phi_0$ is the angle between the nodes ${\bf k}_1$
and ${\bf k}_2$ at the Fermi surface. For the pure $d$-wave symmetry
which we consider here, $\phi_0=\pi/2$, and the integral can be evaluated 
to give
\beq
\lb{Lsingle}
{\cal L}(\epsilon_1, \epsilon_2)=
\frac{1}{\pi} \frac{E_H^2}{(\epsilon_1^2 +\epsilon_2^2)^2},
 \ \ \ \mbox{if $\sqrt{\epsilon_1^2 +\epsilon_2^2}\ge E_H$},
\seq
and zero otherwise. The physical reason for the discontinuity is
that for the nodes at the orthogonal positions 
$\ei^2+\eii^2=E_H^2/\rho^2\ge E_H^2$, so that the 
probability of having the Doppler shifts not satisfying this
inequality is identically zero. In an orthorhombic
system, where the nodes are not at angle $\pi/2$, the 
shape of the distribution is different. 
In analogy with the single node probability
density we can also define the dimensionless energies
$(x,y)=(\epsilon_1,\epsilon_2)/E_H$, and introduce the function
\beq
L(x,y)=E_H^2 {\cal L} (\epsilon_1, \epsilon_2)={1\over\pi (x^2 + y^2)^2},
\ \ \ \mbox{if $x^2+y^2\geq 1$},
\seq
and zero otherwise.

\subsection{Single vortex, ${\bf H}\| \widehat{\bf ab}$}
\label{sec:Pab}

For the magnetic field applied in the superconducting plane 
it has been recently argued that for a relatively
3-dimensional high-T$_c$ material, such as YBCO, the semiclassical approach
still captures the essential features of the quasiparticle behavior.
\cite{Vekhter2} The approach of Ref. \onlinecite{Vekhter2} is to take the
supervelocity field from an  anisotropic London model, but 
to introduce the Doppler
shift only in the dispersion of the quasiparticles with the momenta in
the plane. After rescaling the c-axis to make the unit cell of the 
vortex lattice isotropic, the Doppler shift is given by\cite{Vekhter2}
\beq
\lb{Doppler_ab}
\vsr\cdot {\bf k}_f=\frac{\eab}{\rho}\sin\theta\sin(\phi-\alpha),
\seq
where the angle $\phi$ parameterizes the cylindrical
Fermi surface, 
$\alpha$ is the angle between the direction of the magnetic field
in the  plane and the $x$-axis, and the in-plane energy scale is
$\eab=\eta \eh$, where in the London effective mass
model the anisotropy
$\eta=(\lambda_{ab}/\lambda_c)^{1/2}$.  
In the nodal approximation
(which provides an
excellent agreement with the numerical results \cite{Vekhter2})
the probability distribution of 
the Doppler shift at a single node is given by Eq.(\ref{Psingle})
with $\eh$ replaced by $E_1=\eab |\sin (\pi/4-\alpha)|$ and
$E_2=\eab |\cos (\pi/4-\alpha)|$ respectively for the two pairs of nodes.
Any effects of the three dimensionality reduce the effective value
$\eab$ rather severely \cite{Vekhter2}, so that the estimate obtained 
using the value of $\eta$ for the effective anisotropy in the 
two-dimensional case
can only serve as an upper limit.

For such a geometry the Doppler shifts at the two
neighboring nodes are related by $\epsilon_2=E_2\epsilon_1/E_1$;
in contrast to the case of the field applied along the $c$-axis,
the Doppler shift at one of the nodes uniquely determines the value of the
Doppler shift at the other node independently of the winding angle
$\theta$ in real space. Therefore the two-node probability distribution 
is given by
\beq
\lb{Lsingle_ab}
{\cal L}(\epsilon_1, \epsilon_2)={\cal P}(\epsilon_1)
\delta\biggl(\epsilon_2-\frac{E_2}{E_1}\epsilon_1\biggr),
\seq
and a single average is always sufficient for 
computing the physical quantities in the semiclassical
approximations for the field applied in the 
plane.

\subsection{Vortex solids and liquids}

We now discuss how the probability densities obtained above can be
generalized to the case of vortex solids or liquids. We first consider the
single node probability density $P(x)$. Since this function is normalized, the 
question is what type of the redistribution of the density in Fig.1 one
may expect for realistic vortex structures. As argued above, the high energy
tail of the distribution is entirely determined by the single vortex
physics, and is therefore insensitive to the structure of the vortex state;
the redistribution of weight occurs in the region $x\lesssim 1$ or
$\ep\lesssim \eh$. 

It is also clear that the single vortex picture 
described above underestimates the number of points where the Doppler
shift vanishes. For the supervelocity field of a single vortex 
$|\vsr|> 0$ everywhere in the unit cell, and the Doppler shift vanishes
only for the superfluid velocity direction normal to the nodal directions
in {\bf k}-space. In a vortex lattice there exist points where
$|\vsr|=0$: the high symmetry locations such as midpoints between 
the centers of two neighboring vortices. Consequently, for vortex lattices
$P(0)$ is larger than it is in the single vortex picture. The weight 
shifted to the vanishing Doppler shift, $x=0$,  comes at the price of 
a reduction in the peak in $P(x)$ and moving the
peak  to smaller $x$. The actual shape of the function
depends on the type of the vortex lattice, the number of nearest 
neighbors, and on relative orientation of the basis vectors of 
the vortex lattice with respect to the nodal directions.

As the number of nearest neighbors is increased, so
is the value $P(0)$.
 This value depends not only on the number of zeroes, but also on the
asymptotic behavior of the supervelocity near a point where it
vanishes, $v_s({\bf r}_0)=0$. It is easy to check that
if $v_s({\bf r}-{\bf r}_0)\propto |{\bf r}-{\bf r}_0|^\eta$,
the contribution of this area to $P(0)$ is finite 
for $\eta<2$, is singular but integrable for $2\leq \eta <3$,
and is non-integrable (and therefore non-physical) for $\eta\geq 3$.
In a typical vortex distribution $v_s$ varies linearly with the distance 
from ${\bf r}_0$, so that $P(0)$ remains finite.
We now try to derive analytically an approximate distribution 
which gives a large weight to the probability 
of the vanishing
Doppler shift; we consider it here to model a relatively disordered vortex
state, such as a vortex liquid, and to provide a lower limit of the
magnetic field dependence of the physical quantities.
To make progress we consider a cylindrically symmetric
spatial dependence of the supervelocity, modulated
compared to the single vortex distribution.
Different choices for the modulation of the superfluid velocity are
considered in the literature \cite{Wortis,Tinkham};
in any approach the supervelocity near the vortex core should remain
nearly unmodified  compared to the single vortex velocity field, 
while at the cell boundary $v_s=0$.
Therefore, in the cylindrically symmetric case, the Doppler shift 
(for {\bf H}$\| \widehat{\bf c}$) can be approximated
as 
\beq
\vsr\cdot {\bf k}_f=\eh S(\rho)\sin\theta,
\seq
where $S(\rho\rightarrow 0)\propto 1/\rho$ and
$S(1)=0$. 

Notice that the requirement that $P(0)$ is finite imposes 
restrictions on the decay of $S(\rho)$ as $\rho\rightarrow 0$. 
Since in the cylindrically symmetric model $v_s$ vanishes along
a line rather than at discrete points (as it does for a realistic
vortex distribution), the required asymptotic behavior of $S(\rho)$ 
is different from that of $v_s$ in the system with points of vanishing 
Doppler shift. Nevertheless, as we show below, the appropriate choice of 
$S(\rho)$ allows us to arrive at a probability distribution
close to that obtained by numerical simulations of the vortex liquid.
In such a liquid the distribution is temperature dependent.
A detailed calculation therefore would have  to take into account 
the changes in
the probability density with the temperature in a given material. These 
changes are not well understood beyond simple models, 
and even then are usually accessible only via a
dynamical simulations. We therefore take the point of view that
for a qualitative or semi-quantitative analysis it is sufficient to
consider a model temperature-independent distribution. 
\cite{Vekhter_qc}

Computing the distribution ${\cal P}(\epsilon)$ from 
Eq.(\ref{1dist}) we obtain
\beq
P(x)=\frac{1}{\pi}\int_0^1 \frac{\rho d\rho}{\sqrt{S^2(\rho)-x^2}}.
\seq
Clearly, $P(0)$ is finite when
$S(\rho\rightarrow 1)\propto (1-\rho)^\eta$ with $\eta< 1$.
We use here two different models where the superfluid velocity 
field of a single vortex
is modulated to vanish at the unit cell boundary in 
a fashion which allows analytical progress. 
In the first,
we take 
the modulating factor to be $(1-r^2/R^2)^{1/2}$, 
which leads to 
$S(\rho)=\sqrt{1-\rho^2}/\rho$.
Computing the probability densities as in the previous section we
find
\beq
\lb{Lliq}
{\cal L}(\epsilon_1, \epsilon_2)=
\frac{1}{\pi} \frac{E_H^2}{(\epsilon_1^2 +\epsilon_2^2+E_H^2)^2},
\seq
and
\beq
\lb{Pliq}
{\cal P}(\epsilon)=\frac{1}{2}
\frac{E_H^2}{(E_H^2 +\epsilon^2)^{3/2}}.
\seq
In this case the measure of Doppler shift zeroes is large due to disorder
in the positions of vortices, $P(0)=0.5$. The simplicity of 
this probability distribution makes this choice attractive for further
analytical work.

Another possible choice is  $S(\rho)=\sqrt{1-\rho}/\rho$; it leads to
\beqs
\lb{Lliq1}
&&{\cal L}(\epsilon_1, \epsilon_2)=\frac{1}{\pi}
\frac{E_H^4}{(\ei^2+\eii^2)^3}
\\
\nonumber
&&\qquad\times\Biggl\{1+\frac{\ei^2+\eii^2}{E_H^2}
-\frac{1}{\sqrt{4\frac{\ei^2+\eii^2}{E_H^2}+1}}
\Biggl[ 1+3\frac{\ei^2+\eii^2}{E_H^2}\Biggr] \Biggr\}.
\seqs
Note that as $(\ei^2+\eii^2)E_H^2 \rightarrow 0$ the distribution ${\cal L}$
is finite:
${\cal L}(0,0)= 2/(\pi E_H^2)$. 
The corresponding single node probability
density is given by
\beqs
\nonumber
{\cal P}(\epsilon)=\frac{1}{\pi E_H}
\Biggl\{\Biggr[\frac{E_H^3}{\epsilon^3}+\frac{3 E_H^5}{4\epsilon^5}\Biggr]
&&\arccos\frac{1}{\sqrt{(2\epsilon/E_H)^2+1}}
\\
\lb{Pliq1}
&&-\frac{3E_H^4}{2\epsilon^4} \Biggr\}.
\seqs
For this distribution $P(0)=32/(15\pi)\approx 0.68$, larger than 
the value of $0.5$ given by Eq. (\ref{Pliq}).

The probability density $P(x)$ for all three distributions is shown in 
Fig. 1. In the following we will refer to the distributions 
given by Eqs. (\ref{Lliq})-(\ref{Pliq}) and by Eqs. (\ref{Lliq1})-(\ref{Pliq1})
as liquid I and liquid II respectively. The reason for that
is clear from the inset of Fig.1: these distributions are close to 
those obtained with the help of the Langevin dynamics simulations of
the pancake liquid in Ref.\onlinecite{Vekhter_qc};
as in the vortex liquid they preserve
the cylindrical symmetry of the supervelocity field on average, while
introducing zeroes in that field because of the cancellation
of the supervelocity from neighboring vortices.
For a realistic vortex lattice 
we expect the results for thermodynamic
quantities to be  bracketed by
the values obtained in the single vortex approach, which overestimated the 
effect of the field by under-counting the number of points in
the unit cell of the vortex lattice where the Doppler shift
for quasiparticles near a particular node vanishes, and,
at least approximately, by the liquid II distribution given by
Eqs. (\ref{Lliq1})-(\ref{Pliq1}). The distribution function for the 
pancake liquid can be even sharper peaked at $x=0$, nevertheless we believe 
that the approximate analytic form provides a reasonable low-end estimate for
most experimental situations.
\begin{figure}
\label{figure1}
\epsfxsize=3.3in
\epsfbox{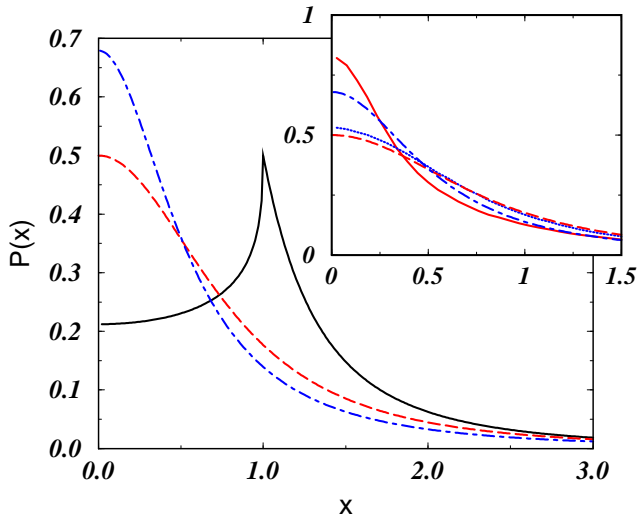}
\caption{Main panel:
probability distribution $P(x)$ in the single vortex
approximation from Eq. (\ref{Psingle}) (solid line), and
for a model vortex liquid states from Eq. (\ref{Pliq}) (dashed line)
for liquid I model, and from Eq. (\ref{Pliq1}) for the liquid II model
(dot-dashed line).
Inset: comparison of the distributions for the model liquid states 
(same notations as in the main panel) with the numerically determined 
distributions for pancake liquid in BSCCO at $T=9$K (narrow distribution)
and $T=67$K (broad distribution) from Ref.48. }
\end{figure}

\subsection{Completely disordered vortex state}

The ``universal'' high Doppler shift behavior of the probability 
density, $P(x)\propto x^{-3}$ results from the strong repulsion 
between the vortex lines which prevents vortex cores from overlapping. 
If the vortices were non-interacting, in a disordered state their 
positions would be completely random, leading to
a Gaussian distribution of the Doppler shifts.
Such an approximation has been 
used by Yu {\it et al.}\cite{Yu} and Franz\cite{Franz1} in their analysis
of the thermal conductivity in the vortex state. Even though it is never 
realized, it is instructive to compare the predictions obtained with such 
a distribution with the results obtained in the framework outlined above.
The comparison may be useful for the extremely anisotropic layered 
superconductors in the geometry with the field applied in
the basal plane. In that arrangement vortices lack proper cores, 
the intervortex repulsion is weakened, and 
we expect significant disorder in vortex positions due to the
presence of defects (such as boundary effects, twin boundaries, etc. ).
Consequently,
the $1/x^3$ asymptotic behavior
does not onset up to large Doppler shifts (very close to the core), and 
over the low (compared to the gap amplitude) energies, 
the probability density decays rapidly.
We therefore also consider in the following the random distribution of
vortices, which leads 
(omitting  factors of $\ln\lambda_L/\xi_0$ in the width of the Gaussian)
to the probability density\cite{Yu,Franz1}
\beq 
P(x)\approx\frac{1}{\sqrt\pi}{\rm e}^{-x^2}.
\seq

We now investigate the dependence of the thermodynamic coefficients on
the magnetic field and the temperature
for different structure of the vortex state and
compare it with the experimentally observed behavior. 

\section{Density of states: pure limit}
\label{sec:dos}

We begin by considering the density of states and the electronic
contribution to the specific heat in the pure limit. While this is one of the 
simplest quantities to analyze, it is the one directly relevant to the
measurements of the field dependence of the specific heat in YBCO single
crystals.\cite{Moler,Phillips,Revaz,Junod} 
To justify ignoring impurities in this analysis we emphasize that
the energy scales associated with the Doppler shift are quite large, and 
at moderate fields exceed the impurity bandwidth even in not too clean
samples, and exceed it by far in the latest single crystals.\cite{Revaz,Junod}
Taking the Fermi velocity $v_f\sim 1.5-2.5\times 10^7$ cm/s \cite{ARPES}, we
obtain $\eh/\sqrt H\sim 30$ K$\cdot$T$^{-1/2}$, and for
YBCO near optimal doping, where $1/\eta\sim 2.5-4$, we obtain
$\eab /\sqrt H \leq 10$ K$\cdot$T$^{-1/2}$, while the impurity bandwidth
$\gamma$ is of the order of a few Kelvin or less.

\subsection{Density of states for ${\bf H}\|\widehat{\bf c}$.}

We first consider the experimental arrangement with ${\bf H}\|\widehat c$.
The density of states in the pure limit is given by 
Eq.(\ref{dospurefield}) leading to
\beqs
\nonumber
N(\omega, H)&=&\int_{-\infty}^{+\infty} d\epsilon N(\omega+\epsilon) 
{\cal P}(\epsilon)
\\
\lb{dosfield}
&=&
\int_{-\infty}^{+\infty}d\epsilon 
\frac{|\omega+\ep|}{\pi v_f v_\Delta}{\cal P}(\epsilon).
\seqs
Introducing the dimensionless variable $x=\ep/E_H$ and considering hereafter
$\w\ge 0$
we find that the
density of states is given by
\beqs
\lb{dos_in_field}
N(\w, H)&=&
\frac{2\omega}{\pi v_f v_\Delta}\int_0^{\omega/E_H} P(x) dx
\\
\nonumber
&&
+\frac{2 E_H}{\pi v_f v_\Delta}\int_{\omega/E_H}^\infty xP(x) dx.
\seqs
The scaling properties of the density of states with
$\w/E_H$ \cite{Simon}
can be made
obvious by rewriting it as
\beqs
\lb{dos_in_field2}
N(\w, H)&=&\frac{E_H}{\pi v_f v_\Delta}F_N\Bigl(\frac{\w}{\eh}\Bigr),
\\
\lb{scaling_N}
F_N(Z)&=&2
\biggl( Z \int_0^Z P(x) dx + \int_Z^\infty xP(x) dx\biggr).
\seqs

The residual density of states at the
Fermi surface is given by
\beq
\lb{dos0}
N(0, H)=M_1\frac{E_H}{\pi v_f v_\Delta}=\frac{M_1}{2v_\Delta}
\sqrt{\frac{H}{\pi\Phi_0}},
\seq
where $M_1$ is the first moment of the probability distribution of the
Doppler shift
\beq
M_1=2\int_0^\infty x P(x) dx,
\seq
which contains all the information about the structure of the vortex state
relevant to the magnitude of the $\sqrt H$ term in the specific heat.
For the probability density given by Eq.(\ref{Psingle}) (single vortex
model)
we then find $M_1^s=4/\pi\approx 1.27$, 
while for the liquid I distribution
given by Eq.(\ref{Pliq}) we obtain $M_1^{l}=1$.  
For liquid II distribution the integral can be evaluated numerically to give
$M_1^{l2}\approx 0.85$, while 
for the completely
disordered distribution of vortices $M_1^g=1/\sqrt\pi\approx 0.56$.
We therefore expect
that $M_1\sim 1$ for any realistic vortex state. Furthermore,
since the number of zeros of the Doppler shift increases
with the increased disorder in the lattice\cite{Vekhter_qc}, we expect
on general grounds that the coefficient 
is larger for the more ordered vortex state. The
residual density of states given by Eq. (\ref{dos0}) is close
to the expression obtained by Won and Maki in
a different approximation scheme \cite{Won}. 

Expanding Eqs.(\ref{dos_in_field2})-(\ref{scaling_N}) 
at low energies $\w\ll\eh$ we find
\beq
\lb{dos_low_w}
N(\w, H)\approx \frac{E_H}{\pi v_f v_\Delta}
\biggl( M_1+\frac{\w^2}{\eh^2} P(0)\biggr).
\seq
Therefore the energy dependence of the density of states
in the field dominated regime is determined
by the probability weight of the vanishing Doppler shift.
As the lattice changes toward a larger coordination number and 
towards disorder, the measure of points where the
superfluid velocity vanishes increases. As a result, the coefficient of
the leading field dependent term, $M_1$, decreases, while the
coefficient of the energy dependent term, $P(0)$, increases.
The values of this coefficient 
are $P^s(0)=2/3\pi\approx 0.21$,
$P^{l}(0)=0.5$, $P^{l2}(0)=0.68$,
and $P^g(0)=1/\sqrt\pi\approx 0.56$ for the single vortex,
liquid, and Gaussian distributions respectively.
It immediately follows that the position
of the crossover from the field dominated
to the zero-field temperature dominated
 behavior in the average density of states is much more sensitive
to the structure of the vortex state than the leading field-dependent
term.

In the effective weak-field range,  $\w\gg\eh$,
the field dependent contribution
is independent of the distribution of
vortices. The vortices are well separated, and the regions where the Doppler 
shift exceeds the temperature are close to the cores, and consequently 
dominated by the universal tails $P(x)=1/2x^3$, yielding
\beq
\lb{weak_field_dos}
N(\w, H)\approx
\frac{\omega}{\pi v_f v_\Delta}
\biggl( 1+ \frac{1}{2}\frac{\eh^2}{\w^2}\biggr).
\seq

The full dependence of the density of states on the energy and the magnetic
field can be obtained from Eqs.(\ref{dos_in_field2})-(\ref{scaling_N}) 
with the probability 
densities discussed above. For the single vortex picture we regain the
result of K\"ubert and Hirschfeld\cite{Kuebert1}
\beqs
\lb{Nsingle}
&&
F^s_N(Z)=\frac{Z}{\pi}
\\
\nonumber
&&\times
\cases{
	\pi \Bigl(1+Z^{-2}/2\Bigr), &
	if $Z\ge 1$;\cr
	Z^{-2}
	\Bigl[ (1+2Z^2)\arcsin Z+ 3Z\sqrt{1-Z^2}\Bigr],
	& if $Z\le 1$.\cr}
\seqs
For the liquid I model we obtain a remarkably simple result
\beqs
\lb{Fliq}
F^{l}_N&=&{\sqrt{Z^2+1}},
\\
\lb{Nliq}
N^l(\w,H)&=&\frac{\sqrt{\w^2+\eh^2}}{\pi v_f v_\Delta}
\seqs
while for 
 the liquid II model the integral can only be evaluated numerically, 
and for 
the Gaussian model
\beq
\lb{N_gauss}
F^g_N(Z)=Z\Phi(Z)+\frac{\exp(-Z^2)}{\sqrt\pi},
\seq
where $\Phi(Z)$ is the probability integral.\cite{GR}
Notice that for the Gaussian distribution the enhancement 
of the density of states in the weak field limit, $\w\gg \eh$ is
vanishingly small in $\w/\eh$, in contrast to Eq.(\ref{weak_field_dos}).
Indeed, in this limit the field-dependent part of the density of states is
determined by the weight in the part of the distribution ${\cal P}(\ep)$ with
$\ep\ge\w$, which is exponentially small.

\begin{figure}
\label{figure2}
\epsfxsize=3.3in
\epsfbox{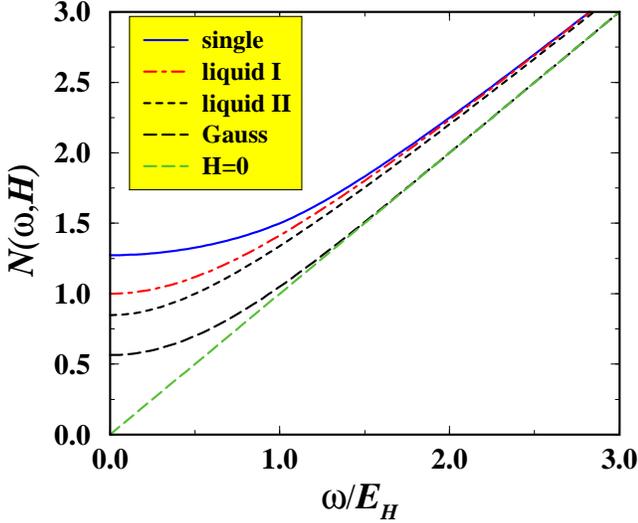}
\caption{Energy
dependence of the density of states in the magnetic field for different 
models of the probability density for the Doppler shift.
Density of states is in units of $E_H/(\pi v_f v_\Delta)$.}
\end{figure}
This difference is clear from Fig.2. The low energy limit
of the density of states depends on the moment of the distribution function, 
and is therefore different for each of the model distributions.
On the other hand the
high energy, or weak field, limit yields the same result for the  
models respecting the asymptotic $x^{-3}$ decay for the probability 
distribution $P(x)$, while the Gaussian model gives the density
of states which is not enhanced relative to the zero-field value.
 The Gaussian model therefore misses the field-dependent 
contribution to the physical quantities at high energies, leading
to incorrect results, especially in the regime $T\leq E_H$.

\subsection{Density of states for ${\bf H}\|\widehat{\bf ab}$.}

We can now analyze in the same framework the anisotropy in
the density of states for the field applied in the superconducting plane, at
an angle $\alpha$ to the x-axis.
As discussed in Section \ref{sec:Pab}, density of states for such
a configuration is a sum over the two inequivalent pairs of nodes,
with 
the different characteristic scales for the 
Doppler shift at each pair of nodes,
\beqs
E_1=\eab |\sin (\pi/4-\alpha)|,
\\
E_2=\eab |\cos (\pi/4-\alpha)|.
\seqs
In the London model
$\eab=\eta \eh$, where $\eta$ is the penetration depth anisotropy
ratio. We emphasize that in reality the value of the ``effective'' anisotropy
depends on the details of the c-axis transport properties \cite{Vekhter2},
and  therefore the estimate of $\eab /\sqrt H \sim 10$ K$\cdot$T$^{-1/2}$
is just an upper limit on its magnitude, and, as we comment below, the 
value inferred from the available experimental data on the specific heat
is lower.

The density of states in the clean limit is given by
\beq
\lb{Nwalpha}
N(\w,H;\alpha)=\frac{1}{2}\Bigl[ N_1(\w,H)+ N_2(\w,H)\Bigr],
\seq
where $N_i$ is computed from Eq.(\ref{dos_in_field2})
as in the previous section but with $E_i$ ($i=1,2$) replacing 
$E_H$. 

As is known \cite{Vekhter2,volovik2}
the residual density of states exhibits
fourfold oscillations as a function of the direction of the applied field
in the plane
\beqs
\lb{dos_fourfold}
N(0,H;\alpha)&=&\frac{M_1}{2}\frac{E_1+E_2}{\pi v_f v_\Delta}
\\
\nonumber
&=&\frac{M_1}{\sqrt 2 \pi}\frac{\eab}{v_f v_\Delta}
\max[|\sin\alpha|, |\cos\alpha|].
\seqs
The minima of the density of states occur when the field is along the
nodal direction, $\alpha=\pi/4+\pi n/2$.
 In that case at two of the four nodes the circulating 
currents are in the plane orthogonal to the direction of the Fermi momentum
at the node, and consequently the Doppler shift vanishes
at all points in real space (either $E_1=0$ or $E_2=0$), as seen in 
Fig.3. In contrast, when the field is along the antinodal direction,
the Doppler shift in non-zero, and all four nodes contribute to the density
of states, leading to a maximum in $N(\w,H)$.\cite{Vekhter2}
\begin{figure}
\label{fig3}
\epsfxsize=3.2in
\epsfbox{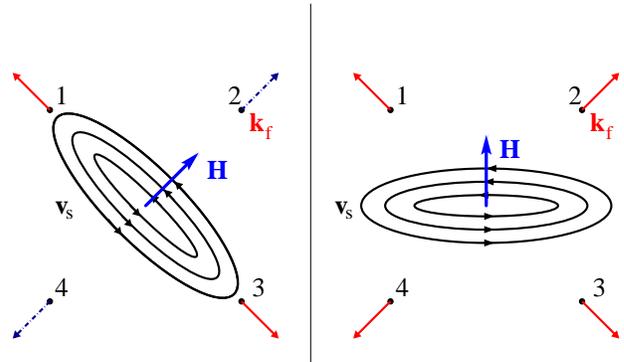}
\caption{Contribution of different nodes to the density of states. 
The nodes are numbered and the direction of the Fermi momentum
is shown at each nodal point.
Left: field along the nodal direction and orthogonal to 
the other pair of nodes. The Fermi momentum
at nodes 2 and 4 (broken line) is orthogonal to the plane 
where supercurrents flow, and the Doppler shift vanishes 
everywhere in space for this pair of nodes. Right: field in the
antinodal direction. The Doppler shift is non-vanishing at some points
in space for each node, and the density of states is maximal.}
\end{figure}
It is important to emphasize that, as is clear from Fig.3, 
it is only for a tetragonal system
that the minima in the density of states occur for the field along
the node. One reason for 
that is that in an orthorhombic system ($m_a\neq m_b$),
for the field applied in a direction other than along the
principal axes of the effective mass tensor,
the directions of the internal and the external
fields differ.\cite{Kogan}
The difference may be quite small in the experimentally relevant
field range; ignoring it, 
Schachinger and Carbotte\cite{Carbotte1} argued that
the minima occur when the field is parallel to the direction
of the Fermi velocity at the node, which differs from the direction towards
the node.

The anisotropy in the density of states given by Eq.(\ref{dos_fourfold})
is $\sim 30$\% for the purely two-dimensional model considered here. Any
three-dimensionality reduces this number severely: if there is a line of 
nodes extending along the $z$-axis, for the field applied towards a
node in the equatorial plane, the Doppler shift
vanishes only for the nodal quasiparticles with momenta
in the plane. For the quasiparticles on the same nodal line but with a
component of the momentum along the $z$-axis the Doppler shift is finite,
consequently the node is still ``active'' in contributing to 
the density of states. In the simplest estimate in a 3-dimensional
system the effect is reduced to 7-8\% \cite{Vekhter2}, in some models 
with a tight binding dispersion along the $c$-axis it may be reduced even
further, to about 4\% \cite{Won2}, making the effect more difficult to detect.

The anisotropy is also rapidly washed out with increased energy
\cite{Vekhter2}.
Since the density of states has a minimum when the field is applied
along a node ($E_1=0$ for example), the corresponding pair of nodes
is ``inactive'' and insensitive to the field; therefore the density of states
increases linearly in energy, as in the absence of a field. For the
field away from the nodal direction the density of states increases
as a square of the energy, see Eq. (\ref{dos_low_w}), resulting in
a rapid suppression of the difference between the two geometries.
For low energies in the limit $\w\ll E_1, E_2$, which can only happen 
if the field
is not close to a nodal direction ($E_1, E_2\neq 0$), we have
\beqs
\lb{dosab_low_w}
N(\w, H;\alpha)&\approx& 
\max[|\sin\alpha|, |\cos\alpha|]
\frac{\eab}{\pi \sqrt 2 v_f v_\Delta}
\\
\nonumber
&\times&
\biggl[{M_1}
+\frac{\omega^2}{\eab^2}\frac{2}{|\cos 2\alpha|}
P(0)\biggr].
\seqs
On the other hand, if $E_1\ll\w\ll E_2$, which may happen when the field is
close to one of the nodes, and $E_1\ll E_2$, we have
\beq
\lb{dosab_node_w}
N(\w, H;\alpha)\approx\frac{1}{2 \pi v_f v_\Delta}
\biggl[M_1 E_2 +\w +\frac{\w^2}{E_2} P(0) +\frac{E_1^2}{2\w}\biggr].
\seq
The anisotropy in the density of states as a function
of the angle for the liquid I model
is shown in Fig.4, at low energies there
is no qualitative difference between the different  models, 
see below. As the energy is increased
the sharp minima fill up, and the resulting anisotropy decreases.
\begin{figure}
\label{fig4}
\epsfxsize=3.2in
\epsfbox{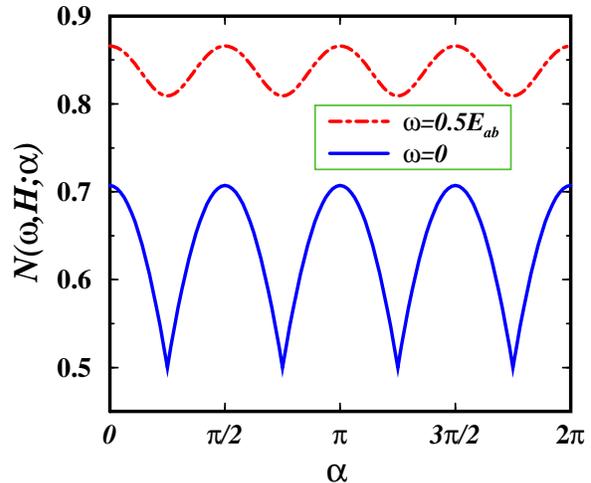}
\caption{Angular dependence of the density of states, measured in
units of $\eab/(\pi v_f v_\Delta)$, on the the direction of
the applied magnetic field. Density of states has been computed with
the model liquid I probability density of the Doppler shift.
Angle $\alpha$ is measured with respect
to the $x$-axis, and the minima are along the position of the nodes.
Notice a significant reduction in the anisotropy at energies 
of the order of $\eab$.}
\end{figure} 

The angular dependence of the density of states vanishes at
higher energies, as for $\w\gg E_1, E_2$ with a realistic 
distribution respecting the asymptotic behavior $P(x)\propto x^{-3}$ for
$x\gg1$
\beqs
N(\w, H;\alpha)&\approx&
\frac{\omega}{\pi v_f v_\Delta}
\Bigl(1+\frac{1}{4} \frac{E_1^2+E_2^2}{\w^2}\Bigr)
\\
\nonumber
&=&
\frac{\omega}{\pi v_f v_\Delta}
\Bigl(1+ \frac{1}{4}\frac{\eab^2}{\w^2}\Bigr).
\seqs
The exact crossover scale 
from the strong to the weak field
regime
depends on the particular choice of the 
probability distribution. This is shown in Fig.5 for the
three different choices of $P(x)$ considered in this work.
\begin{figure}
\label{fig5}
\epsfxsize=3.2in
\epsfbox{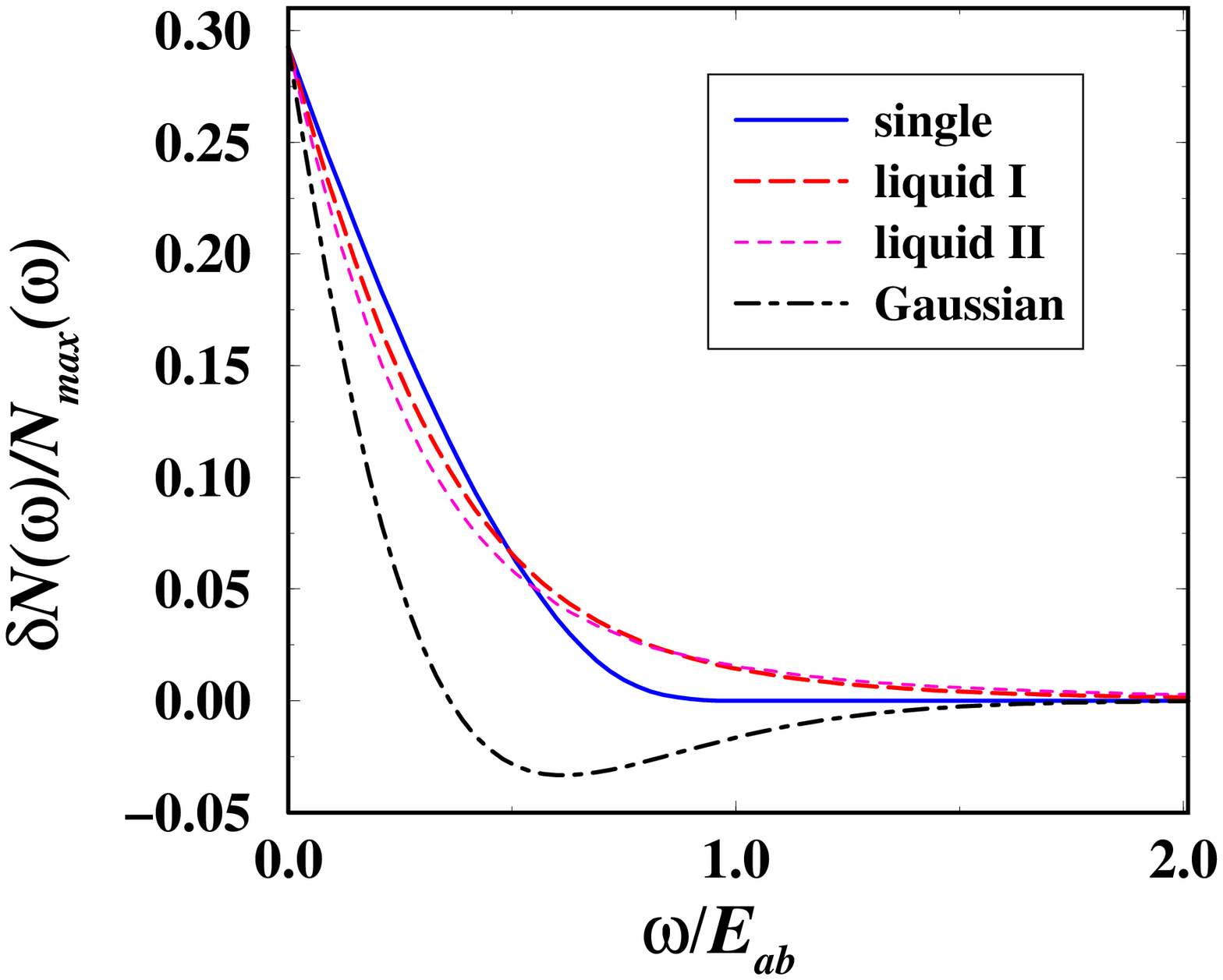}
\caption{Relative anisotropy in the density of states, 
$\delta N(\omega)=N(\w, H; \alpha=0)-N(\w, H; \alpha=\pi/4)$ normalized by the 
maximal $N_{max}(\w)=N(\w, H; 0)$ for different models considered in
the text. Notice that the relative anisotropy at $\w=0$ is identical for
all models, as is clear from Eq. (\ref{dos_fourfold}). In the single vortex 
model the anisotropy vanishes identically at $\w\ge\eab$, see
Eq.(\ref{Nsingle}).
 For the Gaussian
model the exponential asymptotic behavior of the probability distribution
leads to the inverse anisotropy in the intermediate energy range.
The two liquid models yield a very similar dependence of the anisotropy on the
energy.
}
\end{figure}

\subsection{Zeeman splitting}

Typically the Zeeman shift is small compared to the Doppler energy scale, and 
does not modify significantly the density of states.
Indeed, for the field along the $c$-axis,
the spin-up and spin-down density of states
is given by
\beq
N^\pm(\w, H)=\frac{E_H}{\pi v_f v_\Delta}F_N\Bigl(\frac{\w_\pm}{\eh}\Bigr),
\seq
where $\w_\pm=|\w\pm\mu H|$. Therefore, for example,
 the corrections to the residual
density of states due to the paramagnetic contributions 
in the regime $\mu H\ll E_H$ are of the
order
\beq
\frac{\delta N(0,H)}{N(0,H)}\approx
\Bigl(\frac{\mu H}{E_H}\Bigr)^2 \frac{P(0)}{M_1}\ll 1.
\seq
For a quasi three dimensional materials, such as YBCO,
the relevant energy for the field applied in the plane
is $\eab$. 
Even if this energy scale is only 15\% of $E_H$, 
a factor of 2-3 smaller than
estimated, the ratio $\eab/\mu H\sim 6.7/\sqrt H$ T$^{1/2}$ implies a
crossover field of 45 T. Consequently the Zeeman splitting is unimportant 
compared to the Doppler shift
in the experiments performed so far in YBCO.
This is in contrast to more two-dimensional
materials, such as BSCCO, where the response to a parallel magnetic field
is dominated by the Zeeman splitting.\cite{Won3,Sondhi}

The situation is different for 
the geometry with the field applied along a  node;
this was first pointed out in Ref.\onlinecite{Whelan}.
In this case the Doppler shift at one pair of the nodes 
vanishes, and, at $\w=0$, the only contribution to the density of states
at these nodes
is due to the Zeeman splitting. The Zeeman splitting leads to a finite 
contribution to the density of states which is linear in $\mu H$, and
consequently reduces the residual anisotropy
$\delta N(0,H)$. This reduction has been investigated in 
Ref.\onlinecite{Whelan}.

This is however not the only effect of the paramagnetic coupling. 
Since the density of states at the nodes with the vanishing 
Doppler shift is now dominated by the Zeeman splitting at low energies,
the anisotropy is not reduced as rapidly with the increasing energy.
Indeed, if only the paramagnetic effect is take into account,
the total (per particle, i.e. summed over the spins
rather than per spin) density of states is
\beq
\lb{N_Zeeman}
N_Z(\w, H)=\sum_{\alpha=\pm}\frac{|\w+\alpha \mu H|}{\pi v_f v_\Delta}=
\frac{2\max[\mu H, |\w|]}{\pi v_f v_\Delta},
\seq
and therefore the anisotropy increases with $\w$ up to $\w=\mu H$,
where it reaches a maximum.

Let us consider the liquid I model, where
the analytic expression for the density of states is particularly 
simple, the results are not modified substantially if other models are
used. The anisotropy
in the total (summed over the spin directions)
density of states between the nodal and the anti-nodal directions is
given by
\beqs
&&\delta N(\w, H)=\frac{1}{2\pi v_f v_\Delta}
\\
\nonumber
&&\times\Biggl[\sum_{\alpha=\pm}\Bigl(
2\sqrt{\w_\alpha^2+\eab^2/2} - \sqrt{\w_\alpha^2+\eab^2}\Bigr)
-2\max[\mu H, |\w|]\Biggr]
\seqs
As Fig.6 demonstrates, even though the zero-energy anisotropy
is severely reduced upon inclusion of the Zeeman splitting, the 
anisotropy at moderate energies is close to the result obtained
without accounting for the paramagnetic effect. It is clear that the
magnitude of the reduction and the crossover energy depend on
the actual value of $\eab$, and we need a realistic estimate of this
value to evaluate the
impact of the paramagnetic splitting on
the experimental results for the field along a nodal direction.
Such an estimate can be obtained from the analysis of the data on the specific
heat which we discuss in the next section.
\begin{figure}
\label{fig:zeeman}
\epsfxsize=3.2in
\epsfbox{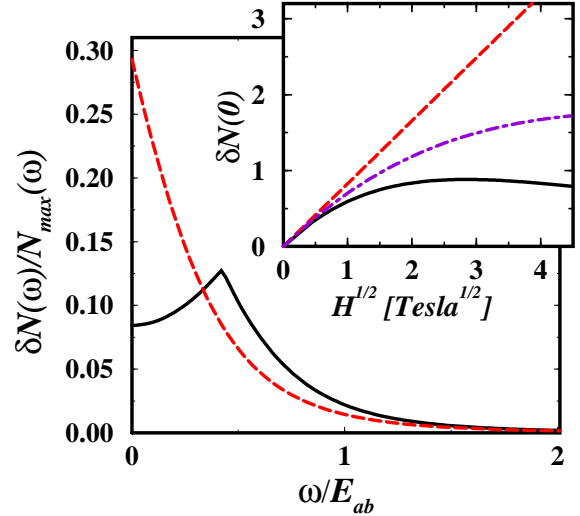}
\caption{Effect of the Zeeman splitting on the anisotropy in
the density of states. Here $\eab/\sqrt H=a$ K$\cdot$T$^{-1/2}$.
Main panel: energy dependence of the anisotropy for the liquid I distribution
with (solid line) and without (dashed line) accounting for the Zeeman
splitting for $a=5$ at $H=10$ tesla.
 Inset: anisotropy in the residual density of states, in
units of $a/2\pi v_f v_\Delta$, as a function of the applied field.
Dashed line: no Zeeman effect, dot-dashed line: $a=10$, solid line:$a=5$.}
\end{figure}

\section{Specific heat and scaling}
\label{sec:spheat}

The information about the density of states is experimentally 
available primarily
via the specific heat measurements, and
 we now address this quantity
in more detail. The preliminary analysis of some of these issues within 
the single vortex picture has been carried out by us before
\cite{Vekhter-pphmf}. Here we concentrate on the effects of different 
distributions, and on the measurability of the specific heat
anisotropy.

\subsection{Specific heat}

The electronic contribution to the specific heat is given by
\cite{Kuebert1,Vekhter-pphmf}
\beqs
\lb{spheat1}
C(T,H)&=&\frac{1}{2}\int_{-\infty}^{+\infty}d\w N(\w, H) 
	\biggl(\frac{\w^2}{T^2}\biggr)
	\cosh^{-2} \frac{\w}{2T}
\\
\nonumber
&=&
T \int_0^{\infty} dx x^2 N(xT, H)\cosh^{-2}\frac{x}{2}.
\seqs
Making use of Eq.(\ref{dos_in_field2}) we can rewrite the specific heat
in the form useful for further analysis. For the field along the $c$-axis,
${\bf H}\| \widehat{\bf c}$, we have
\beq
\lb{spheat}
C(T,H)=\frac{2T\eh}{\pi v_f v_\Delta}\int_0^{\infty} dx x^2 
F_N\Bigl(\frac{xT}{\eh}\Bigr) \cosh^{-2}\frac{x}{2},
\seq
where $F_N$ is given by
Eq.(\ref{scaling_N}).

As a result we find in the limit $\eh\gg T$
\beq
C(T,H)\approx\frac{2T\eh}{\pi v_f v_\Delta}\Bigl(
\frac{\pi^2}{3} M_1 + \frac{7\pi^4}{15} P(0) \frac{T^2}{\eh^2}\Bigr),
\seq
and in the opposite limit, $\eh\ll T$,
\beq
C(T,H)\approx\frac{2T^2}{\pi v_f v_\Delta} \Bigl( 9\zeta(3)
	+\frac{\eh^2}{T^2}\ln 2\Bigr).
\seq

Two quantities which can be compared with experiment are 
the coefficient of the $T^2$ term in absence of the field,
\beq
\lb{c_t}
\gamma_s=\frac{k_B^3}{\hbar^2}\frac{nV_{mol}}{s}
\frac{18\zeta(3)}{\pi v_f v_\Delta},
\seq
and the coefficient of the $T\sqrt H$ term at low temperature
\beq
\lb{c_h}
p=\lim_{T\rightarrow 0}\frac{C(T,H)}{T}=
\frac{\pi^{3/2}}{3}\frac{k_B^2}{\hbar}\frac{n V_{mol}}{s}
\frac{M_1}{v_\Delta\Phi_0^{1/2}}.
\seq
Here $V_{mol}$ is the molar volume,
$s$ is the unit cell size along the $c$-axis, and $n$ is the
number of CuO$_2$ layers per unit cell. 
The presence of both terms has been firmly established
from the analysis of the experimental
data on the specific heat in YBCO \cite{Moler,Phillips,Junod},
however, there remains disagreement about the values of the coefficients
between different groups.

For that material
$V_{mol}=104.6$ cm$^3$/mol, $n=2$, and $s\approx 12$\AA.
The coefficient $p$ can in general be determined to
a higher degree of accuracy, and the values available in the literature
are
$p\approx 0.91$ mJ mol$^{-1}$ K$^{-2}$ T$^{-1/2}$
for moderately clean samples \cite{Moler,Phillips}, and 
more recently obtained $p\approx 1.34$ mJ mol$^{-1}$ K$^{-2}$ T$^{-1/2}$ 
for the ultra-pure single crystals \cite{Junod}.
The analysis of these data in the single vortex picture 
has been carried out by Wang \etal \cite{Junod},
and by Chiao \etal \cite{Chiao2}. 
In that picture the 50\% difference in the coefficient translates into
the same relative difference in the value for the slope of the gap.
In contrast,
according to the previous section, the more
ordered vortex state leads to a larger first moment of the
distribution, and consequently to a larger value of $p$ in Eq.(\ref{c_h}); 
it is therefore
reasonable that a higher quality crystal would have a more ordered
vortex state and hence a larger coefficient $p$. 
If we set $M_1=1$ the  
experimental  
values of $p$ 
lead to the values for the slope of the gap
of $v_\Delta\approx 1.5\times 10^6$cm/s and
$v_\Delta\approx 1.0\times 10^6$cm/s respectively. On the
other hand, taking 
$M_1=4/\pi$ for the pure crystal, 
 yields a larger $v_\Delta\sim 1.27\times 10^6$cm/s, 
leading to a less than 20\% discrepancy between the groups.
The disagreement can be further reduced by assuming a disordered state
with $M_1 <1$ in the ceramic sample of Ref.\onlinecite{Phillips}.
We also note that the pure crystal of 
Ref.\onlinecite{Junod} is overdoped, rather than optimally doped
as in the work of Ref.\onlinecite{Moler,Phillips}, which may contribute
to the  difference in the coefficient.
In combination with the value for the ratio $v_f/v_\Delta\approx 14$
obtained from the universal limit of the thermal conductivity
\cite{Taillefer} this yields
$v_f\sim 1.8\times 10^7$cm/s. This is in reasonable agreement with the
value of the Fermi velocity obtained from the ARPES measurements
in BSCCO,\cite{ARPES} which is believed to have a Fermi surface
similar to that of YBCO.

The coefficient $\gamma_s$ of the temperature dependence has been
measured with significantly larger error bars, and the results
from different groups vary significantly: Moler \etal 
\cite{Moler} reported
the value of 0.1 mJ mol$^{-1}$ K$^{-3}$, Wright and co-workers \cite{Phillips}
obtained $\gamma_s\sim 0.064$ in the same units, while Wang \etal \cite{Junod}
measured 0.21. From the comparison 
with Eq.(\ref{c_t}) we find $v_f v_\Delta\sim a\times 10^{13}$cm$^2$/s$^2$, 
where
$a=2.9, 4.5, 1.4$ for the three values given above. All these yield
the Fermi velocity within a factor of two of the estimate given above. This
implies that in the calculations requiring a cut-off in energy 
the cut-off $E_0\sim 1.3-2.3\times 10^3$ K.

We now turn our attention to the field applied in the $ab$-plane, and 
discuss the specific heat following the general approach of our previous
paper.\cite{Vekhter-pphmf}
The first question which we address is the observability 
of the fourfold oscillations in the density of states. These oscillations
have not been seen in 
the experiments by Moler \etal \cite{Moler}; nor have they
been found in recent measurements on very high quality
single crystals of YBCO.\cite{Junod}
It seems likely that the estimate of $\eab\sim 10$K$\cdot$T$^{-1/2}$
from the purely two-dimensional model is too high, 
and the three-dimensionality
reduces the effect significantly.\cite{Vekhter2} It is also possible
that the orthorhombicity, which shifts the minima in the density of
states away from the $\pi/4$ directions, combined with twinning of 
the crystals used in both experiments reduces the observable anisotropy
significantly.\cite{Carbotte1} However, even in this case, the
in-plane 
anisotropy for the fields of up to 14T used in the experiments by
Wang \etal \cite{Junod}  should be within the 
experimental resolution. A very important observation is that
since the anisotropy in the density of states is washed out rapidly
as the energy is increased, the in-plane anisotropy of  the specific
heat is greatly reduced with increased temperature, \cite{Vekhter2}
as seen in
Fig. 7 the reduction is more rapid for the Doppler shift
density
with the larger weight at low energies.
We only consider here the possible situation when the
configuration of the vortex lattice is identical for the field along
the nodal and the anti-nodal directions; then
the limiting behavior for the specific heat with the field 
along an 
anti-node ($\alpha=0$) and along a node ($\alpha=\pi/4$)
is easily obtained from
Eqs.(\ref{dosab_low_w})-(\ref{dosab_node_w}),
\beqs
\lb{C_an}
C(T,H;0)&=&\frac{\sqrt 2 \eab T}{\pi v_f v_\Delta}
\Bigl[ \frac{\pi^2}{3} M_1 + 
\frac{14\pi^4}{15}\frac{T^2}{\eab^2} P(0)\Bigr],
\\
\lb{C_n}
C(T,H;\frac{\pi}{4})&=&\frac{\eab T}{\pi v_f v_\Delta}
\Bigl[ \frac{\pi^2}{3} M_1 + 9\zeta(3) \frac{T}{\eab}\Bigr].
\seqs
\begin{figure}
\label{fig6}
\epsfxsize=3.2in
\epsfbox{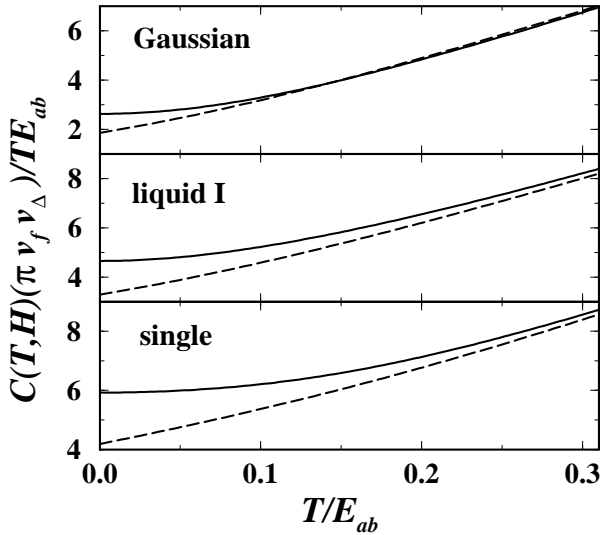}
\caption{Anisotropy in the specific heat between the nodal (dashed line)
and anti-nodal (solid line) directions for different models.}
\end{figure}
Therefore while the amplitude of the $\sqrt H$ term confirms the
estimates for the nodal velocities $v_f$ and $v_\Delta$, and therefore for
the energy scale $E_H$, such a term has not been observed for
the field in the plane, and therefore there is no direct measurement
of $\eab$ available. However, an estimate for this scale can be
obtained from the scaling plots for
the specific heat.

\subsection{Scaling functions}

It has been pointed out by Simon and Lee\cite{Simon} that  on
general grounds the thermodynamic coefficients of
the nodal fermions in a magnetic field should scale with the variable
$T/E_H$; consequently the experimental results can be
interpreted as giving the form of these scaling functions.
The scaling of the specific heat itself follows easily from
Eq.(\ref{dos_in_field2}) for the density of states, and the weak and 
the strong field limits of the scaling function are obtained
from the equations for the specific heat above.
For the field ${\bf H}\|\widehat{\bf c}$, we
define $Z=T/\eh$ and $F_C(Z)={\pi v_f v_\Delta C(T,H)}/(2TE_H)$; then
\beq
\lb{F_C_gen}
F_C(Z)=\int_0^{\infty} dx x^2 
F_N\Bigl(xZ\Bigr) \cosh^{-2}\frac{x}{2},
\seq
with $F_N$ given by Eq.(\ref{scaling_N}). The limits for the 
scaling function follow easily:
\beq
\lb{F_C}
F_C(Z)=\cases{\pi^2 M_1/3 +7\pi^4 P(0) Z^2/15, & if $Z\ll 1$;\cr
		9\zeta(3) Z + Z^{-1}\ln 2,  & if $Z\gg 1$. \cr}
\seq
The numerically determined scaling function is shown in Fig.8.
It is remarkably similar to the scaling plot obtained from
the measured specific heat in Ref.\onlinecite{Phillips}.
In that experiment the crossover scale, 
marking the transition from the field dominated
regime, where $F_C(Z)\approx const$, 
to the temperature dominated regime, has 
been determined
to be $T/\sqrt H \approx 6.5$ K/T$^{1/2}$; a very close value has
been obtained in a more recent experiment of Wang \etal\cite{Junod}. 
As is clearly seen from Fig.8 the value
of the scaling variable at the crossover depends on the structure of 
the vortex state; this is easy to understand from Eq.(\ref{F_C}). The 
zero temperature value of the scaling function is
determined by the first moment of the Doppler shift distribution $M_1$,
while the increase of $F_C$ with the temperature is proportional to
the weight of the distribution at the vanishing Doppler shift, $P(0)$.
Consequently the crossover value, $Z_c$,
can be expected to be proportional to
$\sqrt{M_1/P(0)}$.
As the number of zeroes of the superfluid velocity grows, the weight in
$P(x)$ is shifted towards lower energies, so that $M_1$ decreases while
$P(0)$ increases; these opposing trends
 lead to significant variations in $Z_c$. 
From Fig.8, for the liquid and single vortex models
the crossover occurs around $Z_c\simeq 0.2-0.3$; taking this value as
the experimentally determined crossover point, we arrive
at $ \eh/\sqrt H\sim 30$ K$\cdot$T$^{-1/2}$, in agreement with our
previous estimate. Notice that the crossover occurs at $Z_c\ll 1$;
this is simply the result of a large coefficient of the $Z^2$ term in
the low temperature expansion in Eq. (\ref{F_C}), 
$7\pi^4/15\simeq 45$, while $\pi^2/3\simeq 3$.
\begin{figure}
\label{fig7}
\epsfxsize=3.2in
\epsfbox{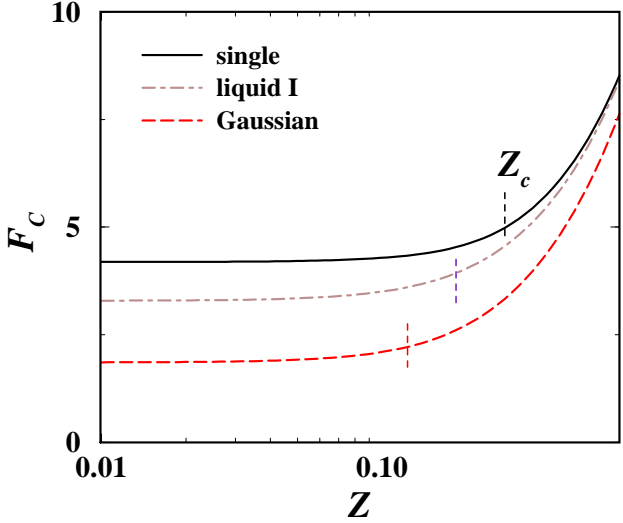}
\caption{Scaling finction for the specific heat. For concreteness
the crossover values $Z_c$
have been defined as the point of a 20\% increase
above the high field flat region: $F_C(Z_c)=1.2 F_C(0)$}
\end{figure}

Similar analysis can be carried out for the field applied in the
plane by introducing $Z_{ab}=T/\eab$ and
$F_{ab}(Z_{ab};\alpha)={\pi v_f v_\Delta C(T,H;\alpha)}/(T\eab)$;
the limiting form of the scaling functions for
$Z_{ab}\ll 1$ can be read off
Eqs.(\ref{C_an})-(\ref{C_n}); at $Z_{ab}\gg 1$ we have
\beq
F_{ab}=18\zeta(3) Z_{ab} + Z_{ab}^{-1}\ln 2.
\seq
The specific heat data of Refs.\onlinecite{Moler,Phillips} 
are analyzed by modeling and subtracting the
``background'' contributions to the specific heat 
(phonons, Schottky anomalies etc.). To avoid the extensive analysis,
Revaz \etal  \cite{Revaz} have looked at the difference between
the specific heat with the field along the $c$-axis, and the field along
the anti-nodal direction, the $c/a-b$ difference
$\delta C(T,H)=C(T,H)-C(T,H;0)$.
The comparison between the results of Ref.\onlinecite{Revaz},
and Refs.\onlinecite{Moler,Phillips} has been a subject of
some controversy, most clearly stated in Ref.\onlinecite{Phillips}.
It has been argued already by the present authors and Carbotte that
the experimental results from these groups are in fact in agreement
\cite{Vekhter-pphmf}, and here we elaborate further on the sources of the
apparent differences. We interpret  $\delta C(T,H)$
as a pure vortex quantity, ignoring the possible elastic contribution
of the vortex lattice and the possible field dependence
of the anisotropy $\eta=\eab/\eh$. The issues raised
in Ref.\onlinecite{Phillips} include the temperature dependence
of $\delta  C(T,H)/T$ in the regime where
$C(T,H)/T$ is essentially insensitive to temperature,
and a form of the scaling function
for $\delta C$ which is quite different from that of $C(T,H)$. 

As is clear from Fig.8 for the field ${\bf H}\|\widehat{\bf c}$
the ratio $C(T,H)/T$ does not depend strongly on the
temperature for $T\lesssim T_H\sim 0.1-0.25\eh$, 
reflecting the energy independence
of the density of states for $\w\ll E_H$. For
the field applied in the plane along the anti-nodal direction
the physics is very similar, up to rescaling of the 
energies, which means that 
the density of states is only constant for $\w\ll \eab\ll\eh$,
and therefore $C(T,H;0)/T$ 
is $T$-dependent above $T_{ab}\sim (0.1-0.25)\eab$.
The difference, $\delta C/T$, becomes temperature dependent at the lower
of the two crossovers, which is at $T\simeq 0.1\eab$, and 
for $T_{ab}\leq T\leq T_H$  it varies with temperature
even though  $C(T,H)/T$ is approximately constant. 

It  is easy to understand the difference in the scaling behavior
between $\delta C$ and $C(T,H)$. Taking the ratio $E_H/\eab=4$
we plot the corresponding scaling functions in Fig.9. 
Even in the regime where the scaling functions $F_C$ is nearly
constant $F_{\delta C}$ is decreasing continuously. 
We therefore believe that there is no contradiction
between the results of Ref.\onlinecite{Junod}, and 
Refs.\onlinecite{Moler,Phillips}. Both the 
temperature dependence of $\delta C/T$ and the difference in the 
behavior of the scaling function reflect the smaller Doppler energy scale
in the plane, $\eab\ll E_H$, and the results of these experiments are,
in fact, quite consistent. Notice that the crossover
in $\delta C$ is much wider than that in $C(T,H)$ because of two
energy scales contributing to it: it extends over a decade in the scaling
variable. Note that in Fig.~9 we have evaluated the specific heat with
the field ${\bf H}\|\widehat c$ in the single vortex approximation, 
while the specific heat for the field along the anti-node in the plane
has been evaluated for the liquid I distribution, to model the expected 
difference in the degree of order in the vortex lattice. Both quantities 
have been evaluated with the single-vortex distribution in a prior 
publication,\cite{Vekhter-pphmf} and there are no qualitative differences
between the two cases.
\begin{figure}
\label{fig8}
\epsfxsize=3.2in
\epsfbox{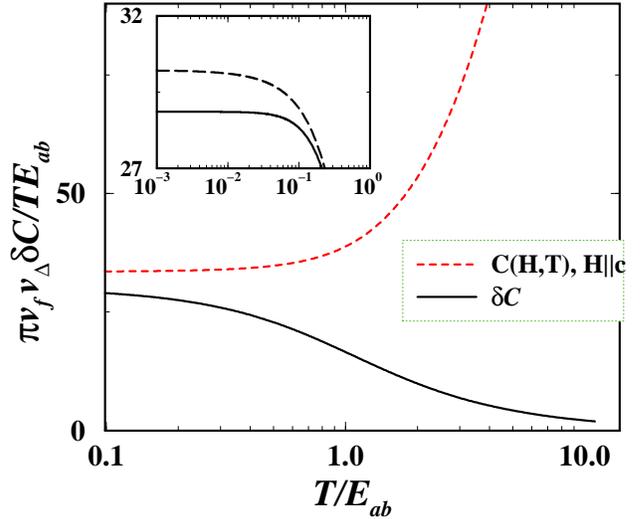}
\caption{Scaling functions for the specific heat with the field applied
along the $c$-axis, $C(T,H)$ and the $c/a-b$ difference 
$\delta C=C(T,H)-C(T,H;0)$.
The former has been evaluated for the single vortex distribution
of the Doppler shift, the latter for the liquid I model
as explained in the text. The behavior
remains essentially unmodified for other forms of the distribution.
Inset: difference between the nodal (top) and anti-nodal directions
disappears on the scale of the larger graph.}
\end{figure}

To further quantify these considerations we note that even though
the crossover to the temperature insensitive
$\delta C$ has not been found in Ref.\onlinecite{Revaz}, the data
suggest that it is close to $T/\sqrt H \sim 0.5$ K T$^{-1/2}$,
which is the lowest value of the scaling variable reached in the paper
(experimental
measurements are limited to the temperatures above $\sim 1.5$K, since 
at lower temperatures the non-vortex contributions to $C(T,H)$ become
dominant). Taking this number as a crossover value of $T/\eab$, 
we estimate $\eab/\sqrt H \sim 3-4.5$ K T$^{-1/2}$; 2-3 times smaller
than the estimate from the London model. 

If the value of $\eab$ is low, it is not surprising
that the in-plane anisotropy between the nodal and the anti-nodal directions
has not been found in the experiments of
Ref.\onlinecite{Junod}: at 14 T, 
$\eab\approx 11-17$ K, and even at the lowest temperature 
where the measurements of Ref.\onlinecite{Junod} have been made
$T/\eab\ge 0.09-0.15$. Then the anisotropy in the density of
states is significantly reduced from the $T=0$ value, see 
the inset of Fig.9.
On the other hand, the data of Ref.\onlinecite{Junod} for the 
field in the plane yield (after the subtraction of 
the Schottky anomaly) a crossover temperature 
between the field dominated and the temperature dominated regimes
close to $T_{cr}\approx 2$ K per T$^{1/2}$.
If this value is taken as corresponding to the 
the crossover in $T/\eab$, it implies a large value of
$\eab/\sqrt H \geq 10-20$ K T$^{-1/2}$. In that case the absence
of the anisotropy can only be explained 
(somewhat unsatisfactorily)
by an appeal to the 
three-dimensionality\cite{Vekhter2} or a combination of the
orthorhombicity
and twinning.\cite{Vekhter2,Carbotte1}
Since part of the experimental difficulty stems from the smallness
of the $\sqrt H$ term with the field in the plane, the analysis 
for that geometry typically 
involves assuming a field dependent contribution 
of that form\cite{Moler,Junod}. However, if $\eab$ is small, the field 
dependence of the specific heat is modified by the Zeeman
splitting, and this splitting has to be taken into account in the analysis.

\subsection{Zeeman splitting}

If the energy scale for the in-plane 
Doppler shift is indeed much smaller than 
the naive estimate from the London model,
$\eab/\sqrt H\sim 3$-$4.5$ K T$^{-1/2}$, 
the Zeeman splitting has a significant effect on the specific
heat with the field applied along a node
in the experimentally relevant range.
The specific heat no longer obeys the scaling properties discussed
above; for the field along a node the contribution of
the Doppler-``inactive'' nodes is given by
\beqs
\lb{zeeman_spheat}
&&C_Z(T,H)=\frac{T^2}{\pi v_f v_\Delta}F_Z\Biggl(\frac{\mu H}{T}\Biggr)
\\
&&F_Z(x)=x
\int_0^x t^2\cosh^{-2}\frac{t}{2} dt
+\int_x^\infty t^3 \cosh^{-2}\frac{t}{2}dt,
\seqs
in agreement with Ref.\onlinecite{Sondhi}, and therefore scales with 
$H$ rather than $\sqrt H$.

As the in-plane anisotropy in the density of states has a maximum for
$\w=\mu H$, the anisotropy in the specific heat also goes through
a maximum; we expect approximately $T_{max}(H)\propto H$.
 We consider here two different cases  
for $a=\eab/\sqrt H$: a large value corresponding to our original
estimate $a=10$K T$^{-1/2}$, and a small value implied by the experiment,
$a=4$K T$^{-1/2}$, and evaluate the specific heat for the liquid I
distribution.
 The main panel of Fig.10 shows the scaling 
plot for the in-plane anisotropy in the specific heat, 
$C_{anis}(T,H)=C(T,H;0)-C(T,H;\pi/4)$ at $H=10$T, so
that the values for the two cases are $\eab\simeq 32$K, and $\eab\simeq 13K$.
 While the anisotropy is 
severely reduced at $T=0$, it becomes close to the values estimated without
accounting for the Zeeman shift at the temperatures $T\simeq 2.2$K and 
$T\simeq 1.8$K
for the two cases respectively, and therefore the anisotropy 
in the experimentally relevant regime is not modified significantly.
Nevertheless, if the absolute magnitude of the anisotropic term
is small, and its field dependence has to be modeled in
the analysis of the experimental results\cite{Junod}, it is important to note
that, as is clear from the inset of  Fig.10, the field 
dependence of the anisotropy is not simply proportional to $\sqrt H$, 
but flattens and decreases at high fields. The deviations are
especially important for small $a$,
since then the maximum of the anisotropy is reached at $H\sim 10$T for
$T=1.5$K, well within the experimental range. We analyze this scenario
in more detail in Fig.11, which demonstrates that
if the coefficient $a$ is small, the maximum in the anisotropy can be 
observed at low temperatures, but moves out of the 
easily accessible range, to $H\ge 20$ T,
at higher $T$. In comparison, if $a\sim 10$, the maximum lies at high
fields for all relevant $T$.
\begin{figure}
\lb{fig:CZ1}
\epsfxsize=3.2in
\epsfbox{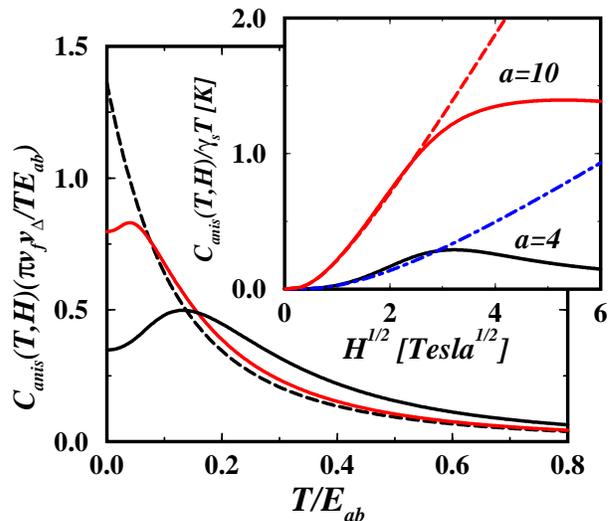}
\caption{Main panel: the anisotropy in the 
in-plane specific heat plotted in the scaling
form at $H=10$ Tesla for the coefficient $a=4, 10$ K T$^{-1/2}$, and without
accounting for the Zeeman shift (dashed line). Inset: The specific heat
at $T=1.5$K, which is close to the lowest experimentally accessible 
temperature, for the same two values of $a$ with (solid line) and 
without (dashed line) accounting for the Zeeman shift.}
\end{figure}
\begin{figure}
\lb{fig:CZ2}
\epsfxsize=3.2in
\epsfbox{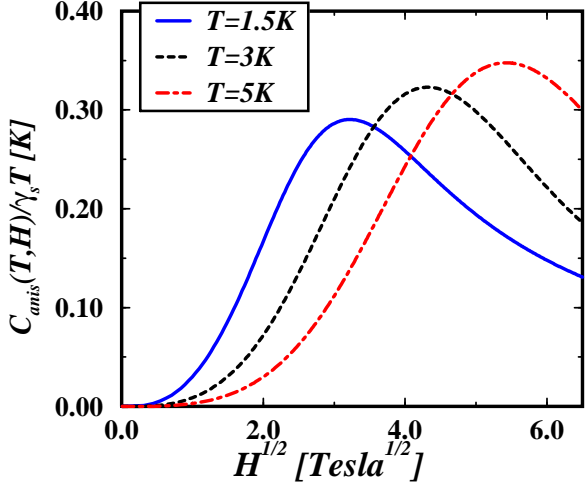}
\caption{Anisotropy o fthe in-plane
specific heat for  $a=4$ K T$^{-1/2}$ as a function of 
the field at different temperatures.}
\end{figure}
Consequently, in the search for the experimental verification 
of the anisotropy in the specific heat, it cannot be assumed
that the anisotropy increases as $\sqrt H$; if the energy scale for
the in-plane Doppler shift is small, the Zeeman splitting modifies the field
dependence of
the specific heat. For the small value $a=4$K T$^{-1/2}$ the maximum
anisotropy, reached in the fields of the order of 10-15 Tesla at $T=1.5$-$3$ K,
is of the order of $0.5$-$0.9\gamma_s$; based on the available 
experimental values for $\gamma_s$\cite{Moler,Phillips,Junod} between
0.064 and 0.21 mJ mol$^{-1}$ K$^{-3}$,
the maximal anisotropy ranges between $0.032$ and
$0.19$ mJ mol$^{-1}$ K$^{-1}$; it is significantly larger for larger
values of $\eab/\sqrt H$.

Recently, Wang \etal attempted to observe  the angular  oscillations we 
have predicted in the
in-plane specific heat.\cite{Junod}   They did not, however,  find appreciable
difference between two measurements with  field applied in the nodal and 
anti-nodal directions. Several reasons
may have contributed: first of all, the YBCO sample used in their experiment
is twinned. Twinning, combined with the orthorhombicity of YBCO is 
expected to reduce the anisotropy.\cite{Vekhter2,Carbotte1}
Here we point out another possible reason for the difficulty in extracting
the difference between the two directions
 from the data: the field dependence of
the anisotropic term is not simply given by  $\sqrt H$, 
as it would be in the absence of the Zeeman term, and as
assumed in Ref.\onlinecite{Junod}. Instead,
the anisotropy increases with the field up to fields of about 10-15 T, and 
decreases thereafter. Consequently, we believe that to confirm the
predicted oscillations experimentally, it is highly desirable to use 
an untwinned crystal, and carry out the measurements at intermediate fields 
(10-15 T) at the lowest possible temperatures, since the anisotropy in the
specific heat is expected to be the largest in this range.

\section{Spin-lattice relaxation rate}
\label{sec:nmr}

We now turn our attention to the calculation of the response functions.
In these calculations the local, in real space, physical quantities
depend on the Doppler shift at both pairs of nodes, and consequently 
the averaging has to be carried out with the two-node probability
density ${\cal L}$ rather than the single node distribution ${\cal P}$.
The simplest example of such a quantity is the average spin-lattice 
relaxation rate which we now consider.
 
Since the NMR measurements on cuprates are typically done in
a magnetic field of $\sim 10$T the effect of the field on the measured 
signal has to be considered in the analysis of the data. There are at least
two effects of the vortex state on the spin-lattice relaxation time. First,
the Doppler shift modifies the local density of states, introducing the
local relaxation rate, which varies from point to point. Second, the
magnetization due to the vortex lattice introduces
inhomogeneities in
the field, leading to the broadening of the resonance line. 
As a result, there are two possible approaches to the analysis. 
In a perfect vortex lattice there exists a one-to-one correspondence
between the local field at a particular point in the unit cell
of the vortex lattice, and the value of the superfluid velocity
at that point. Assuming such a perfect lattice it is therefore
possible to associate the local relaxation rate 
with the relaxation rate at a particular frequency in the resonant line.
Such an approach has been developed theoretically in the semiclassical
framework \cite{Wortis,Morr}, and the results are in qualitative
agreement with the experimental observation that the relaxation rate and
the local density of states are larger in the regions of higher field, i.e.
higher supervelocity \cite{Slichter}.

On the other hand, in a disordered vortex state there is no
unique identification of the local value of the supervelocity corresponding
to a local magnetic field. It may therefore be useful to analyze the
relaxation rate obtained by the ``global'' fit to the resonance line;
especially when the linewidth remains quite narrow in frequency.
It has been shown that the time-decay of the magnetization
is non-exponential as it involves a convolution of many local 
relaxation rates, but that it is possible to describe it with
an effective scattering rate which depends on the field and the temperature.
\cite{Vekhter1,Miami} Usually the analysis of the experimental data is
done assuming a single relaxation rate, and it is therefore important
to understand its behavior in a $d$-wave superconductor.

The analysis of the relaxation time, $T_1$, can be undertaken either by
looking at its magnitude directly, 
or by analyzing the ratio $T_1/\tau_c$, where $\tau_c$ is the relaxation
time at $T=T_c$. The former approach 
involves modeling or estimating from the available data
the matrix element for the interaction; it has been used, for example, in
Ref.\onlinecite{Wortis}.
The latter method is based on making assumptions about the normal state
relaxation in the cuprates. We employ it assuming a normal metallic 
relaxation at $T_c$ with the caveat that this may be only qualitatively
correct for underdoped compounds.

With this assumption for the spin-$\frac{1}{2}$ system 
the magnetization decays
as $m(t)={M(t)/M(0)}=\exp(-t/T_1)$, where the relaxation rate 
in the infinitesimal field is given by
\beqs
\frac{\tau_c T_c}{T_1 T}&=&\int_{-\infty}^{+\infty}d\w
\frac{N^2(\w)}{N_0^2}\Bigl(-\frac{\partial f}{\partial \w}\Bigr)
\\
\nonumber
&=&
\frac{1}{2}\int_0^{+\infty} dx
\frac{N^2(xT)}{N_0^2} \cosh^{-2}x/2,
\seqs
where $N_0=m/2\pi\hbar^2$ is the 2D density of states in the normal state.

In non-zero magnetic field, the decay of the average magnetization is
given by
\beq
\lb{magnetization}
m(t)=4
\int_0^\infty d\epsilon_1 \int_0^\infty 
d\epsilon_2 {\cal L}(\epsilon_1, \epsilon_2)
\exp[-t/T_1(\epsilon_1, \epsilon_2)],
\seq
where the position and Doppler shift dependent 
relaxation rate is determined from
\beq
\frac{\tau_c T_c}{T_1 T}=\frac{1}{2N_0^2}\int_0^{+\infty} dx
N^2(xT,\ei, \eii) \cosh^{-2}x/2.
\seq
The density of states is given by the sum of the contributions of
all nodes, $|\w+\epsilon_i|/(\pi v_f v_\Delta)$ with
($\epsilon_i=\pm\ei,\pm\eii$), which yields
\beqs
\label{dos12}
&&N(\omega, \epsilon_1, \epsilon_2)=\frac{1}{2 \pi v_f v_\Delta}
\\
\nonumber
&&
\times
\cases{ {2\omega}, & if $\omega \geq \max(\epsilon_1,\epsilon_2)$; 
		\cr
	{\omega +\max(\epsilon_1, \epsilon_2)}, & if 
		$\min(\epsilon_1,\epsilon_2) \leq \omega 
		\leq \max(\epsilon_1,\epsilon_2)$;\cr
{\epsilon_2 +\epsilon_1}, & if $\omega 
\leq\min(\epsilon_1,\epsilon_2)  $.\cr}
\seqs
Here again, without loss of generality, we set $\w, \ei, \eii\ge 0$.

The decay 
of the magnetization is therefore non-exponential\cite{Vekhter1,Miami},
as is shown in Fig.12 for a liquid distribution.
In a magnetic field the density of states is enhanced globally,
and therefore the stronger the field, the faster 
the decay of $m(t)$. In the
regime $E_H\gg T$ the density of states is significantly enhanced 
over a large part of the unit cell of the vortex lattice. This high
density of states yields a fast relaxation rate responsible for
the initial decrease in the magnetization. The long-time decay
of $m(t)$ is determined by the slowest relaxation rates, which occur
in the regions  where the superfluid velocity is small and the density
of states is largely determined by the temperature. The two regimes are 
seen in Fig.12 : the large-$t$ tail of $\ln m(t)$
is affected by the field much more  weakly than
the short-time decay.
For all the values of the field it is possible to fit the
time dependence by an exponential, although clearly
the relaxation rate obtained from such a fit differs significantly from
the zero-field rate. We have addressed the fit of the magnetization 
at different time scales in a previous publication.\cite{Vekhter1,Miami}
\begin{figure}
\label{fig:nmr1}
\epsfxsize=3.2in
\epsfbox{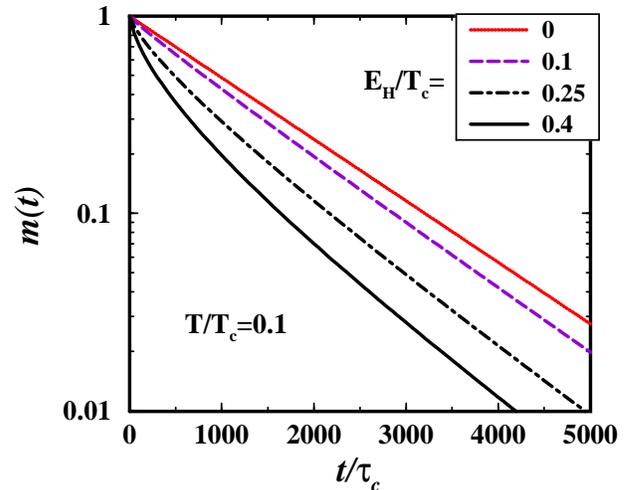}
\caption{Magnetization decay at a fixed temperature for different values of
the magnetic field. We have used 
$2N_0 \pi v_f v_\Delta = k_f v_\Delta \approx 2\D$ (pure $d$-wave), 
and have set
$\D=2.14 T_c$. $m(t)$ has been evaluated for the liquid I model.}
\end{figure}

An important comment concerns the scaling of the magnetization. First of all, 
due to scaling properties of the density of states, the magnetization
decay due to spin-lattice relaxation satisfies
\beq
m(t)=F_m\Bigl(tHT f\Bigl(\frac{T}{\sqrt H}\Bigr)\Bigr),
\seq
where the functions $F_m$ and $f$ can be obtained from the general expression
Eq.(\ref{magnetization}). Moreover, when $\eh/T\gg 1$ the density of states
and therefore the function 
$f$ are nearly constant. Two conclusions follow immediately. First,
at a fixed ratio $T/\sqrt H$ the magnetization depends only on 
the single variable $tTH$. Second, at low
temperatures $m(tTH)$ is independent of 
 the ratio $T/\sqrt H$ at short time scales, when
the relaxation rate is dominated by the field-induced density of
states rather than the temperature-driven density of states. 
The collapse of the low-$T$ data on a single curve as a function of 
$tT$ has been found previously \cite{Guo,Zheng1}, however we are
not aware of an experimental check of such scaling at different fields.
In Fig.13 this behavior is clearly seen.
Deviations from the scaling form are noticeable already for 
$\eh/T\sim 4$, however even in this regime the curves for different
$\eh$ and $T$, but with the
same ratio $\eh/T$, coincide. The scaling is always obeyed
at short time scales, 
where the time decay of the magnetization is determined
by the fast relaxation rates in the regions with a large Doppler shift. On the
other hand, at long time scales the time-dependence
of $m(t)$ is determined by the slowest relaxation rates, in the regions where
the Doppler shift vanishes, and therefore there are always deviations from
the scaling with $tTH$.
\begin{figure}
\label{fig:nmrscale}
\epsfxsize=3.2in
\epsfbox{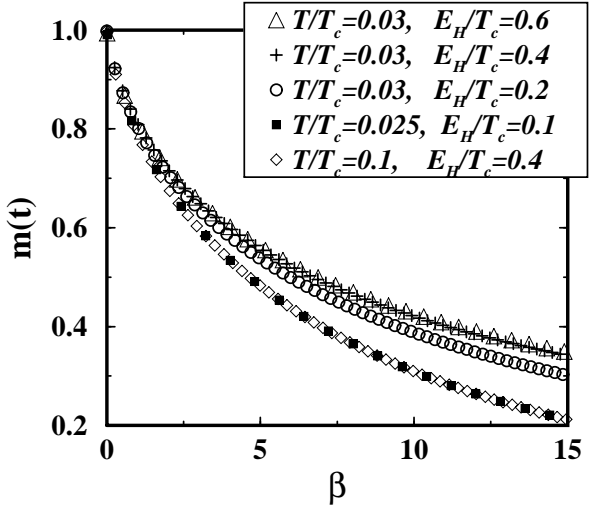}
\caption{Magnetization as a function of 
$\beta=tT\eh^2/\tau_c T_c\D^2$. The ratio $T/\eh$ is identical for
the bottom two sets.}
\end{figure}

Since the  
long time scale decay is determined by the measure of the points with
small Doppler shift, it depends crucially on the probability
density ${\cal L}(\ei, \eii)$. In particular, there is a dramatic difference
between the single vortex picture, where $\ei^2+\eii^2\ge \eh^2$, and
the lattice or liquid states, where this restriction is lifted: magnetization
decays much faster in the single vortex picture, as can be seen in
Fig.14. Notice that for the very early times, when the
magnetization decay is determined by the regions with the highest
Doppler shift, the two distributions give the same result. Therefore the 
effective relaxation rate obtained from the exponential fit
depends not only on the field but also on the structure of the vortex state.
\begin{figure}
\label{fig:nmr2}
\epsfxsize=3.2in
\epsfbox{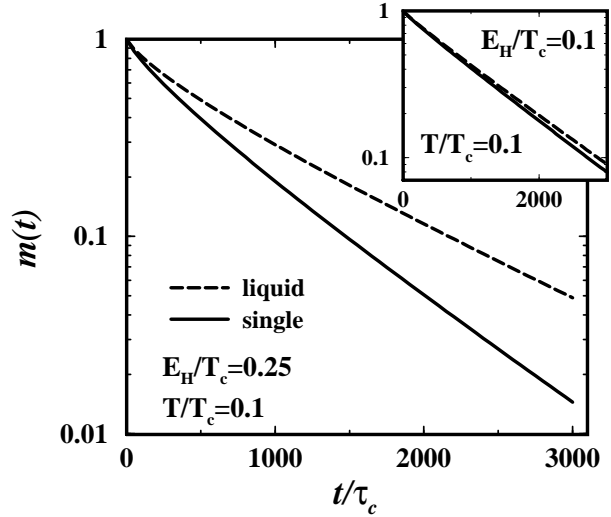}
\caption{The difference between the liquid (liquid I)
and the single vortex models 
in a strong (main panel) and weak (inset) magnetic field. Parameters are
the same as in Fig. 12.}
\end{figure}
The difference in the behavior for the 
two types of the vortex state can be understood
from the analysis of the magnetization. Noticing that 
${\cal L}(\ei, \eii)={\cal L}_1 (\ei^2+\eii^2)$ due to symmetry
( the probability density should be even in both $\ei$ and $\eii$, 
and should be symmetric under the interchange
$\ei\leftrightarrow\eii$), and introducing 
polar coordinates $x=(\ei^2+\eii^2)/\eh^2$, $\tan\phi=\ei/\eii$, we arrive at
\beqs
m(t)&\approx&
\int_{x_0}^\infty dx {\cal L}_1 (x) e^{-\beta x/4} I(\beta x/4),
\\
I(z)&=& \int_0^\pi \exp(-z\sin\phi) d\phi,
\seqs
where $x_0\approx T^2/\eh^2$, and
$\beta=tT\eh^2/\tau_c T_c \D^2$.
This form shows explicitly that there is an approximate scaling
with the variable $\beta$, and that the scaling is obeyed better
the smaller the ratio $T/\eh$.

Due to the exponential the integral over $x$ is cut off
at $\beta x\gg 1$, so that we only need to evaluate $I(z)$
for $z\sim 1$. In that case all angles $\phi$ contribute to the
integral, and $I(z)\approx \pi \exp(-bz)$ with $b\sim 1$ 
(in contrast, $I(z)\simeq 2/z$ for $z\gg 1$), leading to
\beq
m(t)\approx\pi\int_{x_0}^\infty dx {\cal L}_1 (x) e^{-\beta s x/4},
\seq
with $s=1+a\sim 1$.
Consequently, the long time scale limits ($\beta\gg 1$)
for the single vortex and liquid I distribution respectively
\beqs
m(t)&=&\int_1^\infty dx \frac{e^{-\beta s x/4}}{x^2}
\sim \frac{\exp(-\beta s/4)}{s\beta/4},
\\
m(t)&=&\int_{x_0}^\infty dx \frac{e^{-\beta s x/4}}{(x+1)^2}
\sim \frac{\exp(-\beta s x_0/4)}{s\beta/4}.
\seqs
As a result, the long time decay for the liquid regime is governed
by the relaxation rate close
to the $\beta x_0\propto T^3$ behavior expected for $H=0$, 
while for the single vortex
model the relaxation rate is proportional to $\beta\propto TH$.

In reality however the decay of $m(t)$ at long time scales
is usually not measured, and at intermediate times the
detailed analysis of the time dependence of the magnetization 
taking into account the non-exponential form of $m(t)$
is complex. It is possible to define an effective relaxation
rate, however, the weight of the components
of the magnetization with fast and slow decay is different
for different definitions, and the resulting effective
relaxation rate is different, as we now illustrate.
One possible approach is to define the effective rate as
\beq
\lb{t1eff}
\frac{1}{T_1^{eff}}=4
\int_0^\infty d\epsilon_1 \int_0^\infty 
d\epsilon_2 {\cal L}(\epsilon_1, \epsilon_2)
\frac{1}{T_1(\epsilon_1, \epsilon_2)}.
\seq
Unlike the average for the magnetization, Eq.(\ref{magnetization}),
which has the largest contribution from the slowest relaxation rates,
in this averaging procedure the weight of short relaxation rates is high, and 
a cutoff of the energy integral near the core is required. 
To leading order in $\ln E_0/\eh$ the relaxation rate in the
field dominated regime is then given by
\beq
\lb{teff1}
\frac{1}{T_1^{eff}}\approx\frac{\pi+2}{2\pi}\frac{1}{\tau_c}
\frac{T}{T_c}\frac{\eh^2}{\D^2}\ln\frac{E_0}{\eh}.
\seq
The relaxation rate given by this expression
is expected to overestimate the rate of the decay of the magnetization.
Fast relaxation occurs near the cores, where the effective field is higher,
and therefore in the component of the signal away from the original position
of the resonance line.\cite{Wortis}
Alternatively, we can define the average relaxation {\it time}
\beqs
\lb{t2eff}
{\tau_1^{eff}}&=&\int_0^\infty m(t)dt
\\
\nonumber
&=&
4
\int_0^\infty d\epsilon_1 \int_0^\infty 
d\epsilon_2 {\cal L}(\epsilon_1, \epsilon_2)
{T_1(\epsilon_1, \epsilon_2)}.
\seqs
This average has a large contribution from slow relaxation rates, 
and we expect the effective rate to be underestimates since
over experimentally relevant time scales the slowest rates do not
contribute to the magnetization decay appreciably.
Indeed, for the cases of the single vortex and the liquid distributions 
we obtain in the field dominated regime
\beqs
\frac{1}{\tau_1^{eff}}&\approx&\frac{\pi}{8}\frac{1}{\tau_c}
\frac{T}{T_c}\frac{\eh^2}{\D^2}, \ \ \ \ \ \ \ \ \ \
\mbox{single vortex},
\\
\frac{1}{\tau_1^{eff}}&\approx&\frac{\pi}{32}\frac{1}{\tau_c}
\frac{T}{T_c}\frac{\eh^2}{\D^2}\Bigl[\ln\frac{E_H}{T}\Bigr]^{-1}, \ \ \ \ \ 
\mbox{liquid I}.
\seqs
The coefficient 
in the last expression is significantly smaller than the expression
given by Eq.(\ref{teff1}).
We can now compare this expression with the result of Ref.\onlinecite{Guo},
where it was found that $\tau_c T_c/T_1 T\approx 0.2$ at $H=11$ T at low $T$. 
From our estimate of $\eh$ it follows that at this field
$\eh\sim 100$K. Taking $E_0\simeq 1500$ K, we find this ratio for the 
average rate to be
be $0.35$, 
while the average relaxation time procedure yields the values of 
$0.06$ and $0.005$ (at $T\simeq 5$ K) respectively. The experimental value
is between the two estimates, as expected.

\section{Impurity scattering}
\label{sec:impurity}

\subsection{The self-energy}

In the presence of impurity scattering the frequency is renormalized
according to $\wt=\omega-\Sigma(\wt)$. The self energy $\Sigma(\wt)$
depends on the momentum integral of the Green's function, and therefore
on the Doppler shifts at all nodes. Consequently, in all the calculations 
involving the impurities, the local quantities depend on both $\ei$ and
$\eii$. 

Here we consider the impurity scattering in the unitarity limit.
The strategy for the calculation is 
as follows. The self energy is given by Eq.(\ref{Sigma}); to evaluate
this expression in a field we have to introduce the Doppler shift in 
the Green's function as before, and solve for the self energy 
self-consistently at each node. In other words, there is a distinct Doppler
shift at each node, and the self-consistency requires that the scattering
to other nodes with their respective Doppler shifts be taken into
account self-consistently. Therefore we can write
\beq
\lb{SigmaH}
\Sigma(\wt, \ei , \eii)=-{n_i}[G_0(\wt, \ei , \eii)]^{-1},
\seq
where, from Eq. (\ref{G11}),
\beqs
\lb{G0field}
G_0(\wt, \ei , \eii)&=&-\frac{1}{2\pi}\sum_{\alpha=\pm\atop n=1,2}
\frac{\omega_{\alpha,n}+i\wii}{v_f v_\Delta}
\\
\nonumber
&\times&
\Biggl[
\ln\frac{E_0}{\sqrt{\omega_{\alpha,n}^2+\wii^2}} +
i \arctan\frac{\omega_{\alpha,n}}{\wii}\Biggr],
\seqs
and
\beqs
\omega_{\alpha,n}&=&\w+\alpha\epsilon_n - {\rm Re} \Sigma(\wt, \ei , \eii),
\\
\wii&=& -{\rm Im} \Sigma(\wt, \ei , \eii).
\seqs

We now focus our attention on
the cases of weak $\w\ll\eh\ll \gamma$, and strong
$\w\ll \gamma\ll\eh$ fields at low temperatures. Setting $\w=0$
it is clear immediately that the real part
of the momentum integral of the Green's function at each node in 
Eq. (\ref{G0field}) is odd in the Doppler shift, $\epsilon_n$, and 
therefore vanishes upon summation.\cite{Kuebert1} 
As a result, the self energy has only
the imaginary part given by Eq.(\ref{SigmaH}) with
\beqs
G_0(\wt, \ei , \eii)&=&-\frac{i}{\pi}
\frac{\wii}{v_f v_\Delta}
\ln\frac{E_0^2}{\sqrt{\ei^2+\wii^2}\sqrt{\eii^2+\wii^2}}
\\
\nonumber
&-&
\frac{i}{\pi}\frac{\ei}{v_f v_\Delta}
 \arctan\frac{\ei}{\wii}
-\frac{i}{\pi}\frac{\eii}{v_f v_\Delta}
 \arctan\frac{\eii}{\wii}
\seqs
In the strong field limit, $\wii\ll\ei,\eii$, we obtain
for the the density of states
\beq
N(\ei, \eii)=-\frac{1}{\pi}{\rm Im}G_0(\wt, \ei , \eii)=
\frac{1}{2\pi}\frac{|\ei|+|\eii|}{v_f v_\Delta},
\seq
as expected, see Eq.(\ref{dos12}).
The quasiparticle damping in this regime is given by
\beq
\wii\approx\frac{2n_i v_f v_\Delta}{|\ei|+|\eii|}
\sim\frac{\gamma^2}{|\ei|+|\eii|}\ll\gamma.
\seq
In the weak field impurity dominated regime, $\wii\gg\ei,\eii$,
on the other hand,
mv the field-induced change in the density of states is quadratic 
in the Doppler shift and is given by
\beq
\delta N(\ei, \eii)\approx \frac{1}{4\pi^2v_f v_\Delta}
	\frac{\ei^2+\eii^2}{\gamma},
\seq
where $\gamma$ is the zero energy scattering rate
which has been defined in Section \ref{sec:semiclassics}.
Then to the leading order
in $\ln E_0/\eh$ the average change in the density of states is given by
\beq
\delta N_s(0, H)\approx\frac{E_H^2}{2\pi^2\gamma v_f v_\Delta}
\ln\frac{E_0}{\eh}\propto H\ln\frac{H_0}{H}
\seq
for the  single vortex and the liquid distributions.
For the Gaussian model
the change in the density of states is smaller by
a logarithm of a large number, 
$\delta N_s(0, H)/\delta N_G(0,H)=2\ln E_0/\eh$. This behavior is illustrated 
in Fig.15. Notice that the three distributions 
yield different high-field slopes, corresponding to the different
values of the moment $M_1$.
\begin{figure}
\label{fig:dirtydos}
\epsfxsize=3.2in
\epsfbox{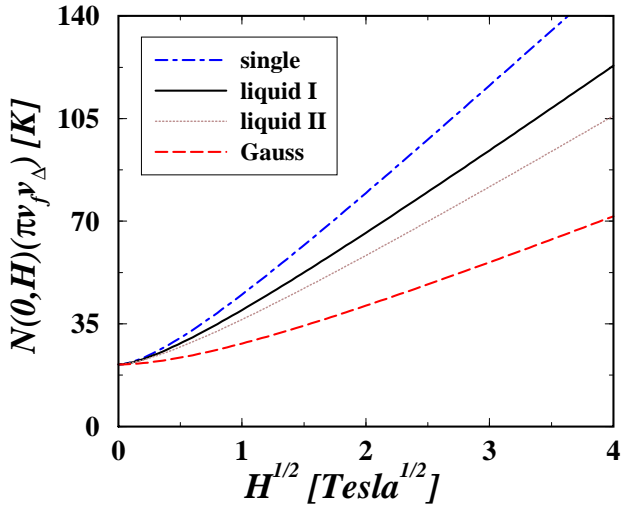}
\caption{Residual density of states as a function of the field. The impurity
scattering is taken in the unitarity limit, with 
 $(n_i\pi v_f v_\Delta)^{1/2}=20$ K. The parameters are $E_0=1500$K,
 $E_H/\sqrt H=30$ K T$^{-1/2}$.}
\end{figure} 
These results are in agreement with the previous work\cite{Kuebert1,Barash}.
The low energy scattering rate, $\gamma$, provides the new energy scale
in addition to the average Doppler shift $E_H$ and the temperature $T$.
At low temperatures, the competition between $E_H$ and $\gamma$ 
determines the behavior of the density of states, and 
in the field dominated regime, $E_H\gg \gamma$, the
density of states strongly 
depends on the probability density of the Doppler shift, as it does 
in the pure limit. 
 The dependence of the self energy on the magnetic 
field is crucially important for the analysis of the transport
properties in the vortex state, and we will discuss these
issues in detail in  a separate paper.

\section{Conclusions}

In this paper we have discussed the semiclassical approach to the vortex
state of unconventional superconductors and have applied it to
the analysis of the thermodynamic properties
of a two dimensional $d$-wave superconductor, which we take as a model for
the high-T$_c$ cuprates at low energies.
Our main point is that within the semiclassical approach the 
dependence of the measured quantities on the magnetic field is
sensitive to the structure of the vortex state and the distribution
of the supercurrents.
This is shown in an  approach 
which
involves introducing the Doppler shift due to circulating
supercurrents into the quasiparticle dispersion, and computing
the physically measured magnetic field dependent
quantities as a spatial average of their local
values in the vortex state. The major step which has enabled us to 
move beyond the standard single vortex description is the
rewriting of the spatial average in terms of the average over a probability
density of the Doppler shift at a particular node or at a pair of nodes.
We have analytically computed these probability densities for
the single vortex picture, for model liquid distributions, and 
for a non-physical, albeit often used, completely random
distribution. 
We have argued that this approach is easily applicable to
any given distribution of vortices, and that the single vortex and the liquid 
models typically give the upper and the lower limits of the field
dependence since they over- and under-estimate respectively
the number of the points in the vortex lattice unit cell where
the Doppler shift dominates the physical picture.

We have applied this approach to the analysis of the electronic specific 
heat in the vortex state and to the description of the spin-lattice
relaxation of the NMR magnetization.
In the former case the specific heat depends on the single-node
probability distribution. The values for the Fermi velocity at
the nodes, as well as of the slope of the superconducting gap, 
determined from such an analysis are consistent with the
values inferred from other experimental measurements, and the values directly
determined from the photoemission. Moreover, noticing that 
the magnitude of the $\sqrt H$ term is larger
for the more ordered vortex state, has allowed us to reduce the
discrepancy between the results for the gap slope obtained by different 
experimental groups. We have also emphasized that the difference 
in the form of the scaling function obtained by these groups is
naturally explained as a consequence of the smaller Doppler shift
energy scale for the field applied in the plane; this work confirms our
earlier assessment on the basis of the single vortex picture.
\cite{Vekhter-pphmf}

Since the analysis of the scaling plots allowed us to estimate the
energy scale of the Doppler shift for the field in the plane, and since this
energy scale is smaller than the London model estimate of 
our previous work\cite{Vekhter2}, we have investigated here whether
the anisotropy in the specific heat between the experimental arrangements
with the field applied along a node and between the two nodes is
observable. We have paid special attention to the effect of the 
Zeeman splitting, which becomes more important for 
smaller in-plane scale
$\eab$. We have found that, while the zero temperature anisotropy
is significantly reduced compared to the case of no Zeeman splitting, as 
predicted\cite{Whelan}, the anisotropy does not decrease with the temperature 
and in the experimentally relevant temperature range
the magnitude
of the anisotropy is weakly affected by the inclusion of the Zeeman 
splitting.
The field dependence of the anisotropic specific heat may however
be modified quite significantly, and this change has to be taken into
account when analysing the experimental data.

We have considered the spin-lattice relaxation rate as an example of
a response function which depends on the probability distribution
at two nodes; in contrast to the specific heat the contributions of 
the nodes are not simply additive. It has been known that
the magnetization decay is non-exponential due to a distribution
of the local relaxation times.\cite{Vekhter1} We have shown here that
the effective relaxation rate obtained from a fit to an exponential at short
or long time scales is different, and at long times depends crucially on the 
structure of the vortex state. We have predicted a scaling form of the
magnetization, and discussed the existing evidence for such a scaling; 
of course more experimental work checking this prediction would be
highly desirable. We have also introduced an effective relaxation rate
obtained from a global fit
to the magnetization, and found that it agrees qualitatively with the 
available
experimental results.

In general, the structure of the vortex state may change quite dramatically
as the temperature and the applied field are varied; one example of
such a change is the melting of the vortex lattice. In such a situation
we expect a change in the spatially averaged thermodynamic quantities measured
in experiment which reflects the transition from one type of
distribution to another. Moreover, as the degree of ordering in the vortex 
lattice depends on the history of the sample, the measured field dependence
varies accordingly.
For example, the coefficient of the $\sqrt H$ term in the 
electronic specific heat should, in general,  depend on whether the sample
has been cooled in an applied field or in zero field: in the latter case
the vortex state is more disordered. Whether these effects are observable 
experimentally depends crucially on the quality of the sample since
the changes may be rather small, nevertheless, in clean untwinned samples 
they may be measurable.

Finally, to illustrate that in the presence of impurities the 
two-node probability density is always  required we have analysed the
density of states in the impurity-dominated regime for different 
structures of the vortex state. This part of our work will be developed
further in
the analysis of the transport properties, which warrants a separate paper.

\acknowledgements

It is a pleasure to acknowledge discussions and corresondence with
L. N. Bulaevskii and A. E. Koshelev.
This research has been supported in part by 
the Department of Energy under contract W-7405-ENG-36 (I.~V.), by
the NSERC of Canada
and the Cottrell Scholarship program of Research Corporation (E.J.N.)
and by the NSF through Grant No. DMR-9974396 (P.J.H.). I.V. acknowledges 
Aspen Center for Physics for hospitality during 
the early stages of this work. P.~J.~H. is grateful for
hospitality and support of ITP Santa Barbara via Grant No. 
NSF-PHY-94-07194.

\end{document}